\documentclass[twoside, openright, titlepage, fleqn, headinclude, 12pt, a4paper, BCOR5mm, footinclude]{scrbook}

\usepackage[utf8]{inputenc}
\usepackage[T1]{fontenc}
\usepackage[english]{babel}
\usepackage[square, numbers]{natbib}
\usepackage[fleqn]{amsmath}
\usepackage{packages/dia-classicthesis-ldpkg}
\usepackage[eulerchapternumbers, subfig, beramono, eulermath, parts, subfigure, dottedtoc]{packages/classicthesis}
\usepackage{xfrac}
\usepackage{marginnote}
\usepackage{graphicx} 
\usepackage{float}
\usepackage{setspace}
\usepackage{amsmath}
\usepackage{amssymb}
\usepackage[lined,boxed,commentsnumbered]{algorithm2e}
\newcommand{\myTitle}{Modello nonlineare di un sistema audio e suo inverso\xspace}
\newcommand{\mySubtitle}{Nonlinear model and its inverse of an audio system\xspace}
\newcommand{\myDegree}{Corso di Laurea Magistrale in Informatica\xspace}
\newcommand{\myName}{Alessandro Loriga\xspace}
\newcommand{\myProf}{Prof. Gregorio Landi\xspace}
\newcommand{\mySupervisor}{Dott. Daniele Bernardini\xspace}
\newcommand{\myFaculty}{Scuola di Scienze Matematiche, Fisiche e Naturali\xspace}

\newcommand{\myUni}{\protect{Università degli Studi di Firenze}\xspace}

\newcommand{\myTime}{Anno Accademico 2015-2016\xspace}

\usepackage{geometry}
\renewcommand*{\lstlistingname}{Source Code}
\newcommand*{\noaddvspace}{\renewcommand*{\addvspace}[1]{}}
\addtocontents{lol}{\protect\noaddvspace}
\let\oldlstlistoflistings\lstlistoflistings
\renewcommand{\lstlistoflistings}{%
    \begingroup%
        \let\oldnumberline\numberline%
        \renewcommand{\numberline}{\!\!\!\!\!\!\!\!\!\lstlistingname~\oldnumberline}%
        \oldlstlistoflistings%
    \endgroup
}
\colorlet{punct}{red!60!black}
\definecolor{background}{HTML}{EEEEEE}
\definecolor{delim}{RGB}{20,105,176}
\colorlet{numb}{magenta!60!black}

\usepackage{perpage}
\MakePerPage{footnote}

\hypersetup{%
    colorlinks=true, linktocpage=true, pdfstartpage=3, pdfstartview=FitV,%
    colorlinks=false, linktocpage=false, pdfstartpage=3, pdfstartview=FitV, pdfborder={0 0 0},%
    breaklinks=true, pdfpagemode=UseNone, pageanchor=true, pdfpagemode=UseOutlines,%
    plainpages=false, bookmarksnumbered, bookmarksopen=true, bookmarksopenlevel=1,%
    hypertexnames=true, pdfhighlight=/O,%nesting=true,%frenchlinks,%
    urlcolor=webbrown, linkcolor=RoyalBlue, citecolor=webgreen, %pagecolor=RoyalBlue,%
    pdftitle={\myTitle},%
    pdfauthor={\textcopyright\ \myName, \myUni, \myFaculty},%
    pdfsubject={},%
    pdfkeywords={},%
    pdfcreator={pdfLaTeX},%
    pdfproducer={LaTeX with hyperref and classicthesis}%
}

\graphicspath{{img/}}

\AtBeginDocument{\renewcommand{\thepart}{\Roman{part}}}

\titleformat{\chapter}[display]%
	{\relax}{\mbox{}\oldmarginpar{\vspace*{-3\baselineskip}\makebox[40pt][r]{\color{halfgray}\chapterNumber\thechapter}}}{-5pt}%
	{\raggedright\spacedallcaps}[\normalsize\vspace*{.8\baselineskip}\titlerule]%
	
\renewcommand{\backrefnotcitedstring}{\relax}%(Not cited.) 
\renewcommand{\backrefcitedsinglestring}[1]{(Citato a pagina~#1.)} 
\renewcommand{\backrefcitedmultistring}[1]{(Citato alle pagine~#1.)}

\newtheorem{theorem}{Theorem}

\lstdefinelanguage{json}{
    basicstyle=\normalfont\ttfamily,
    numbers=left,
    numberstyle=\scriptsize,
    stepnumber=1,
    numbersep=8pt,
    showstringspaces=false,
    breaklines=true,
    backgroundcolor=\color{background},
    literate=
     *{0}{{{\color{numb}0}}}{1}
      {1}{{{\color{numb}1}}}{1}
      {2}{{{\color{numb}2}}}{1}
      {3}{{{\color{numb}3}}}{1}
      {4}{{{\color{numb}4}}}{1}
      {5}{{{\color{numb}5}}}{1}
      {6}{{{\color{numb}6}}}{1}
      {7}{{{\color{numb}7}}}{1}
      {8}{{{\color{numb}8}}}{1}
      {9}{{{\color{numb}9}}}{1}
      {:}{{{\color{punct}{:}}}}{1}
      {,}{{{\color{punct}{,}}}}{1}
      {\{}{{{\color{delim}{\{}}}}{1}
      {\}}{{{\color{delim}{\}}}}}{1}
      {[}{{{\color{delim}{[}}}}{1}
      {]}{{{\color{delim}{]}}}}{1},
}

\begin{document}

	\frenchspacing
	\raggedbottom
	\pagenumbering{roman}
	\pagestyle{plain}
	
%********************************************************************
% Front matter
%********************************************************************

	%--------------------------------------------------------------
% title.tex (tesi.tex main file)
%--------------------------------------------------------------
\begin{titlepage}
	\begin{center}
		\large
		\hfill
		\vfill
		\begingroup
			\spacedallcaps{\myUni}\\
			\myFaculty \\
			\myDegree \\ 
			\vspace{0.5cm}
			\includegraphics[scale=.065]{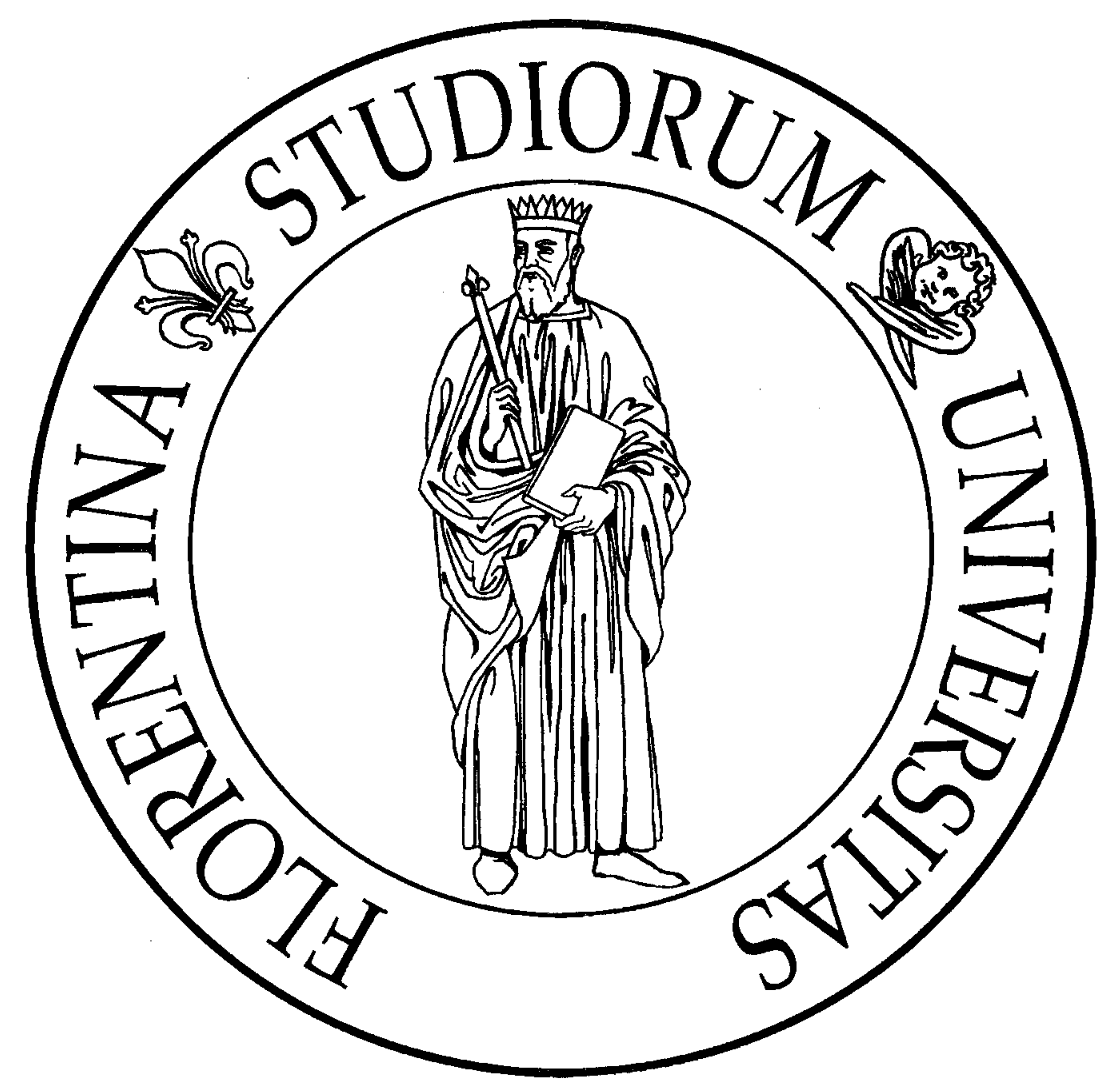}\\
			\vspace{0.5cm}
			Tesi di Laurea
		\endgroup
		\vfill
		\begingroup
			\color{Maroon}\spacedallcaps{\myTitle}\\
			\spacedlowsmallcaps{\mySubtitle}\\
			\bigskip
		\endgroup
		\spacedlowsmallcaps{\myName}
		\vfill
		Relatore: \textit{\myProf}\\
		Correlatore: \textit{\mySupervisor}
		\vfill
		\myTime
		\vfill
	\end{center}
\end{titlepage}
%--------------------------------------------------------------
% back titlepage
%--------------------------------------------------------------
\newpage
	\thispagestyle{empty}
	\hfill
	\vfill
	\noindent\myName: \textit{\myTitle: \mySubtitle,} \myDegree, \textcopyright\ \myTime
%--------------------------------------------------------------
% back titlepage end
%--------------------------------------------------------------
	
	\pagestyle{scrheadings}
	\refstepcounter{dummy}
        \pdfbookmark[1]{\contentsname}{tableofcontents}
        \setcounter{tocdepth}{2} % <-- 2 includes up to subsections in the ToC
        \setcounter{secnumdepth}{3} % <-- 3 numbers up to subsubsections
        \manualmark
        \markboth{\spacedlowsmallcaps{\contentsname}}{\spacedlowsmallcaps{\contentsname}}
        \tableofcontents 
        \automark[section]{chapter}
        \renewcommand{\chaptermark}[1]{\markboth{\spacedlowsmallcaps{#1}}{\spacedlowsmallcaps{#1}}}
        \renewcommand{\sectionmark}[1]{\markright{\thesection\enspace\spacedlowsmallcaps{#1}}}
%	\input{front/abstract}
%	\input{front/aknowledgements}
	
%********************************************************************
% Body matter
%********************************************************************
	
	\pagenumbering{arabic}
	\cleardoublepage

\pdfbookmark[1]{Ringraziamenti}{ringraziamenti}

%\begin{flushright}{\slshape    
%	We have seen that computer programming is an art,\\ 
%	because it applies accumulated knowledge to the world,\\ 
%	because it requires skill and ingenuity, and especially\\
%	because it produces objects of beauty.}\\
%	\medskip --- Donald E. Knuth
%\end{flushright}
%
%\bigskip

\begingroup

	\let\clearpage\relax
	\let\cleardoublepage\relax
	\let\cleardoublepage\relax
	
\chapter*{Aknowledgements}
I am using this opportunity to express my gratitude to everyone who supported me throughout the course of my studies. I am thankful to Maria Carmela, she always encouraged me to follow my dreams. Thanks to my parents and my sister for the continuous support. I am sincerely grateful to all my friends for sharing their lifes with me, in particular Ranieri, he is like a brother to me.\\\\
I express my warm thanks to Prof. Gregorio Landi and Daniele Bernardini for their help and guidance. \\\\
I would also like to thank Daniele Battaglia, Leonardo Tassini, Elisabeth Dumont, Jessica Hole and all the people who provided me help, advice or simply support.\\\\
Thank you all,\\
Alessandro

\endgroup
	\part{Introduction and problem definition}
\chapter{Introduction}
In the last years a large interest has been taken in High Fidelity audio. A very important factor for this expansion is connected to improvements in the software and hardware tools. In particular, thanks to the continuous evolution of the DSP processor, it's becoming possible to apply a large number of corrections to the signal in real-time.\\\\
The common implementations of audio fidelity improvement are obtained as linear analysis and filters, which will be presented in the next chapter. These methodologies can be applied in very simple way with high performance. In effect, performance has represented the most significant barrier for audio application.\\
Many implementations of new technologies have regarded \textit{noise reduction}. The principle of this method is simple and well known; it is based on the fact that if we sum a wave to a disturbing wave with the same frequency and form, but of opposite phase, a cancellation effect can be obtained. Therefore, if the noise wave is known, it can be deleted from the final output.\\
This is simple to obtain in laboratory, but it is very difficult to handle in a context where  the noise is not known a priori. For this reason, important investments in research have been made in this field and a dedicated methodology has been defined: the \textbf{Active Noise Control}(ANC). Interesting utilization of the ANC can be found on \cite{reviewANC} or in \cite{ANCneural} and part of this thesis will discuss this topic. \\
This project employs another technique, called \textbf{Adaptive Filters}, which is also applied to correct the signal. However, this does not exclude the use of the ANC in future developments to correct the echo noise.\\\\
This thesis is structured as follows: the first two chapters are devoted to introducing the instruments used in following.
In chapter three, we show a very rudimentary model in order to understand the large number of the problems associated with this context.\\ 
In chapter four, a standard method is presented to handle weakly nonlinear system. This model has been theoretically analyzed very well during the last 50 years, even though it is used in practical applications only with the computational evolution of the last years. In fact only a few systems on the market employ this method.\\
Particular attention will be paid to the results of the test.In conclusion, we will explain the overall development of this project.

\chapter{Introductory definitions and preliminary equations}

\section{Signals and systems}
\paragraph{Signals}
With the term of \textit{signal} will be defined a function depending by one or more variables containing information of the physic state.\\
This definition is general and signals of very different nature can be defined. For an image, we can consider the information contained in it as a signal, in particular as function of two variables that represent the value of the pixels respect the spatial coordinates $(x,y)$. For a video, this is a function of three variable, the spatial coordinates and the time.\\
In the case of the sound, it's a function of the time. \\\\
Several different signal properties can be expressed for our signals:
\begin{itemize}
    \item continuous-time
    \item discrete-time
    \item continuous amplitude
    \item discrete amplitude
    \item periodic
    \item non-periodic
    \item finite energy
    \item finite power
\end{itemize}
A signal is \textit{continuous-time} if it assumes a value for every instant of time. $\exists x(t),\forall t$. If a signal is defined only for a set of discrete values of $t$, it's called a \textit{discrate-time} signal.\\ 
A similar definition can be expressed for the continuous amplitude and discrete amplitude, considering the values assumed by the function.\\
A signal $x(t)$ is called \textit{periodic} if exists a constant $T$ (period), such that:
\[ x(t) = x(t + kT), \forall t, \forall k \in I\]
Otherwise the signal is non-periodic. Another important relation related to the periodic signals is the \textit{frequency}; the inverse of the period $T$:
\[ f=\frac{1}{T} \]
The dimension of the frequency is the hertz (Hz), that is $s^{-1}$. So a signal with a frequency of $80$Hz, has a period of $\frac{1}{80}$ second, i.e. every $\frac{1}{80}$ second, the same values are repeated.\\
The \textit{energy} of a signal $x(t)$, as far the electric currents, is:
\begin{equation}
E=\int_{-\infty}^{\infty} |x(t)|^2 dt
\end{equation}
Note that the dimension of the energy is dependent from the dimension of $x$. The signal has \textit{finite energy} if:
\[ E=\int_{-\infty}^{\infty} |x(t)|^2 dt < \infty \]
In addition we will define the \textit{mean power} as:
\begin{equation}
P = \frac{1}{T} \int_{-\frac{T}{2}}^{\frac{T}{2}} |x(t)|^2 dt
\end{equation}
A signal $x(t)$ has \textit{finite power} if:
\[ P = \lim_{T\to \infty} \frac{1}{T} \int_{-\frac{T}{2}}^{\frac{T}{2}} |x(t)|^2 dt < \infty \]
A \textit{finite} signal will be a signal defined over a finite time interval. More specifically, we will use finite discrete-time periodic signals.
\paragraph{Systems}
We will also use the concept of systems, a system is a function that maps a signal into another signal.\\ 
Formally, consider a system $H$ that transforms input signal $x$ into an output signal $y$. So, let:
\[ x = [D \rightarrow R] \]
\[ y = [D' \rightarrow R'] \]
Then
\[ H = [D \rightarrow R] \rightarrow [D' \rightarrow R'] \]
This systems are often and also in this case, visualized as component blocks that have connections between them. So, in this sense, we can talk about systems and interconnection in an abstract way without the worry of how is made this transfer function. We will work with \textit{black box}, i.e., systems that work in terms of its inputs and outputs without any knowledge of its internal workings.\\
In this work we will use the \textbf{time-invariant} systems, it means that an output does not depend explicitly on time. This kind of behavior can be expressed with this property:\\\\
\textit{
Let $t$ the variable that indicates the time, if for the input signal $x(t)$ the system produces an output $y(t)$ and for any time shifted input, $x(t + \delta)$, results in a time-shifted output $y(t + \delta)$; then the system is time-invariant.}\\\\
Each system have a particular characteristic function $h$ called \textbf{impulse response}, that describes how an input $x$ is transformed in an output $y$. A linear or nonlinear impulse response will be considered. The most used and studied systems are the linear time-invariant systems (\textit{LTI}), they will be explained in the next chapter.\\
Another important characteristic of the system is the causality:\\\\
\textbf{Definition}: Let $h$ the impulse response of a system $H$, $H$ is \textbf{causal} if and only if:
\[ h(t) = 0, \forall t<0 \]
otherwise it is non-causal.\\
This means that the output depends only from the past and the current input of the system, not from the future. All realizable systems in physical applications are causal systems.\\\\
\begin{figure}[h]\centering
\includegraphics[scale=.3]{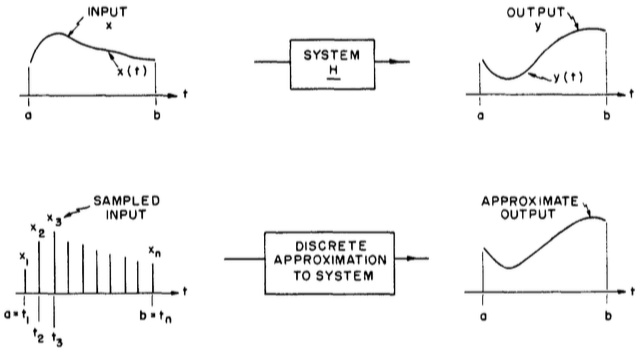} 
\caption{Approximation of a system at discrete time.} \label{fig:nonlinearoperator}
\end{figure}
The mathematical representation of a system is an operator, which is defined as a function whose independent and dependent variables are themselves functions of time.
Considering an operator $H$,  corresponds to the system shown in fig. \ref{fig:nonlinearoperator}, it maps the input functions of time $x$, into the output functions of time $y$.
\[ y = H(x) \]
We will use the common notations to indicate the operator, without the parentheses. So, to write the transformation from $x$ to $y$ through $H$, we will write:  \[y=Hx\]
\section{Shannon Sampling Theorem}
\begin{figure}[H]\centering
\includegraphics[scale=.5]{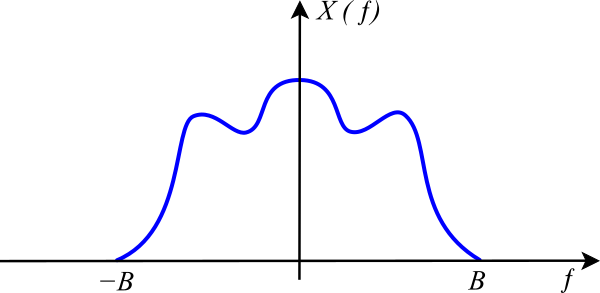} 
\caption{Spectrum of a bandlimited signal} \label{fig:bandlimited}
\end{figure}

\begin{figure}[H]\centering
\includegraphics[scale=.4]{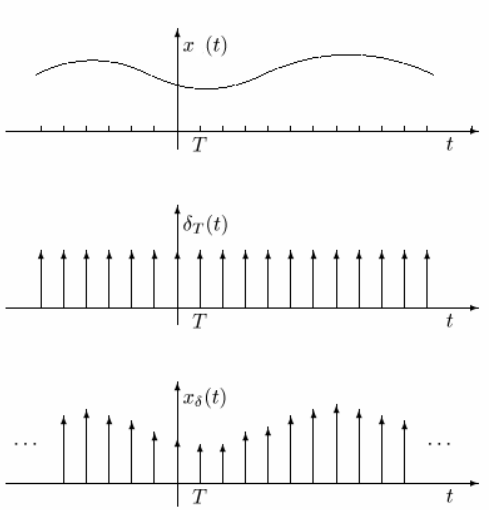} 
\caption{Uniform sampling} \label{fig:samplingdelta}
\end{figure}

The sampling theory is a fundamental concept in digital signal processing. Its role is very important because it allow to connect a continuous-time signal with a discrete-time one, without loss of information.\\ 
In particular, we refer to a continuous-time \textit{band-limited} signals, i.e. a signal with a Fourier transform that is zero above a certain threshold, denoted with $B$ in fig \ref{fig:bandlimited}.
The transformation of a continuous-time signal to a discrete-time signal can be done with and sampling process, the samples are equally spaced along the time axis. This can be modeled exploiting the \textit{impulse signal} $\delta$ \cite{oppenheim}, defined as:
\begin{equation}
\delta (n) = \begin{cases} 0 &\mbox{} n \neq 0 \\ 
1 & \mbox{} n=0 \end{cases}
\label{eq:deltadirac}
\end{equation}
The discrete-time signal with sampling period $T$ can be expressed as:
\begin{equation}
x_{\delta}(n) = x(n) * \delta_T (n)
\label{eq:convolutionXwithDelta}
\end{equation}
Where
\[\delta_T(k) = \sum_{n=-\infty}^{\infty} \delta (k-nT) \]
The scheme is in fig. \ref{fig:samplingdelta}
\textbf{Shannon Sampling Theorem}:
\begin{theorem}
If a function $f(t)$ does not contain frequencies higher than $W$ cps (cycle per second, Hz), it is completely determined by giving its ordinates at a series of points spaced $\frac{1}{2W}$ seconds apart.
\end{theorem}
The sampling theorem asserts that a band-limited signal, observed over the entire time axis, can be perfectly reconstructed from its equally spaced samples taken at a rate which exceeds twice the highest frequency present in the signal.\\
Related to this, we introduce also the the \textit{Nyquist frequency} that is half of the sampling rate of a discrete signal, it is used many times in practical applications.\\\\
The sampling theorem was first introduced in information theory by C. E. Shannon in 1940. However, this theorem has been published before by several authors including E. T. Whittaker, H. Nyquist, J. M. Whittaker, V. A. Kotel’nikov, and its historical roots have been often discussed. Many extensions and generalizations of the Shannon sampling theorem exist but for our purpose we will use only the simplest sentence.\\\\
However, in many practical applications, the signal is not strictly band-limited and it is observed only over a finite time interval. Sometimes the sampled values are not exactly known if some noise occurs in acquiring the samples. These differences from the ideal scenario influence the accuracy of the signal reconstruction.\\
When the assumptions of the Shannon sampling theorem are violated, a special type of error called \textit{Aliasing} is inserted in the signal reconstruction. The frequency components of the original signal that are higher than half the sampling rate are folded into lower frequencies.

\section{Fourier series}
A large use will be done of the Fourier series and Fourier transform, in particular Fourier transform will be frequently used to analyze the signal.\\
A \textbf{Fourier series} is an expansion of a periodic function $f(x)$ in terms of an infinite sum of sines and cosines. In particular the Fourier series of a periodic function $f(x)$ of interval $[-\pi,\pi]$ is given by:
\begin{equation}
 f(x) = \frac{1}{2}a_0 + \sum_{n=1}^\infty a_n cos(nx) + \sum_{n=1}^\infty b_n sin(nx),
 \label{eq:fourierpi}
\end{equation}
where
\[ a_0 = \frac{1}{\pi} \int_{-\pi}^\pi f(x) dx \]
\[ a_n = \frac{1}{\pi} \int_{-\pi}^\pi f(x)cos(nx) dx \]
\[ b_n = \frac{1}{\pi} \int_{-\pi}^\pi f(x)sin(nx) dx \]
For a function $f(x)$ periodic on an interval $[-L,L]$ instead of $[-\pi,\pi]$, a simple change of variables can be used to transform the interval of integration from $[-\pi,\pi]$, as in equation (\ref{eq:fourierpi}) to $[-L,L]$. \\
Let
\[ x=\frac{\pi x'}{L} \]
\[ dx=\frac{\pi dx'}{L} \]
We can solve for $x'$ gives $x'=Lx/\pi$
\begin{equation}
 f(x') = \frac{1}{2}a_0 + \sum_{n=1}^\infty a_n cos(\frac{n\pi x'}{L}) + \sum_{n=1}^\infty b_n sin(\frac{n\pi x'}{L}),
\end{equation}
where,
\[ a_0 = \frac{1}{L} \int_{-L}^L f(x') dx' \]
\[ a_n = \frac{1}{L} \int_{-L}^L f(x')cos(\frac{n\pi x'}{L}) dx' \]
\[ b_n = \frac{1}{L} \int_{-L}^L f(x')sin(\frac{n\pi x'}{L}) dx' \]

\begin{figure}[h]\centering
\includegraphics[scale=.7]{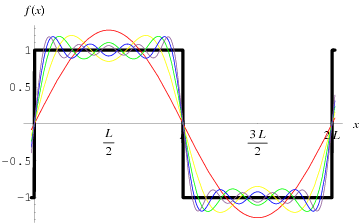} 
\caption{Fourier approximation for periodic functions.} \label{fig:fourierseries}
\end{figure}
We can see an example of the approximation of the functions using a finite Fourier series in fig \ref{fig:fourierseries}.
\subsection{Gibbs phenomenon}
\begin{figure}[h]\centering
\includegraphics[scale=.8]{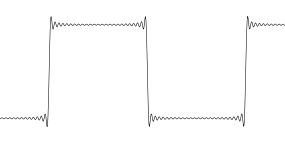} 
\caption{Gibbs phenomenon near the discontinuities} \label{fig:gibbs}
\end{figure}

The \textit{Gibbs} phenomenon is a fluctuation in the Fourier series occurring at simple discontinuities.
The $n$-th partial sum of the Fourier series has large oscillations near the jump. The overshoot does not tend to zero as $n$ increases, but approaches a finite limit.\\
In practical applications, this behavior creates a particular error near the discontinuity during the approximation of the function, this error consists in a large oscillation of the signal near the discontinuity (figure \ref{fig:fourierseries} or \ref{fig:gibbs}).\\\\
Considering the frequency response of a filter, the error produced by the Gibbs phenomenon in the transition band is a ripple in the pass-band (rp). Ripple in the stop-band (rp) is used for the part of the error referred to the stop-band transfer function.

\subsection{Fourier transform}
\begin{figure}[h]
\begin{minipage}{.5\textwidth}
\includegraphics[scale=.3]{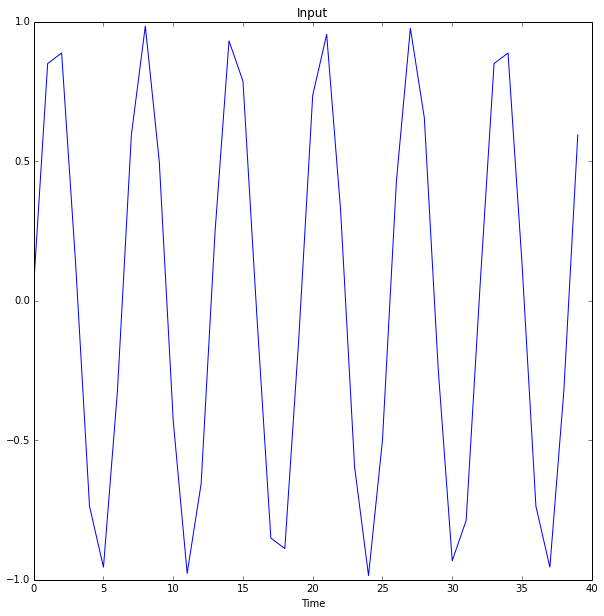} 
\end{minipage}
\begin{minipage}{.5\textwidth}
\includegraphics[scale=.3]{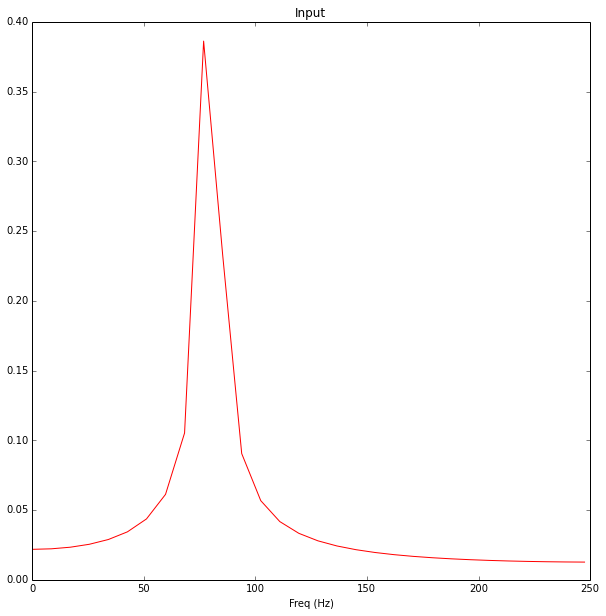} 
\end{minipage}
\caption{Discrete time-domain signal and its spectrum obtained with DFT, the sampled points of the functions are linked with a line to show the wave form.}\label{fig:transform80}
\end{figure}

Fourier transform is a very useful tool used to analyzed a function in the frequency domain. This transform is a generalization of the Fourier series in complex domain. Let $f(x)$ the function we call 
Fourier transform $\mathcal{F}[f(x)]$:
\begin{equation}
F(k)	=	\int_{-\infty}^\infty f(x)e^{-2\pi i k x}dx
\end{equation}
We can define the \textit{inverse} of this transform $\mathcal{F}^{-1}[F(k)]$ with:
\begin{equation}
 f(x) =	\int_{-\infty}^\infty F(k)e^{2\pi i k x}dk
\end{equation}
An important property of the Fourier transform is the linearity. If $f(x)$ and $g(x)$ have Fourier transforms $F(k)$ and $G(k)$, and $a,b$ are scalars, then
\[\int_{-\infty}^\infty [af(x)+bg(x)]e^{-2\pi i k x}dx=\]
\begin{equation}
= a \int_{-\infty}^\infty f(x)e^{-2\pi i k x}dx + b \int_{-\infty}^\infty g(x)e^{-2\pi i k x}dx = aF(k) + bG(k)
\end{equation}
similarly for the inverse. The Fourier transform is symmetric:
\[ F(k) = \mathcal{F}[f(x)](k) \Rightarrow F(-k) = \mathcal{F}[f(-x)](k) \]
For digital applications we need to use a discrete domain, so let $f(t_k)$, denoted with $f_k$, the sampled value in $t_k$, $k=0,...,N-1$. Formally the discretization step can be expressed using the Dirac $\delta$ function:
\[ f(t) = \sum_{n=-\infty}^{\infty} f_n \delta (t-n\tau)\]
where $\tau$ is the sampling period.\\\\
The \textit{Discrete Fourier Transform} (DFT) $F_n = \mathcal{F}_k[\{f_k\}_{k=0}^{N-1}](n)$ is defined as:
\begin{equation}
 F_n = \sum_{k=0}^{N-1} f_k e^{-2 \pi i n k/N}
\end{equation}
And its inverse $f_k = \mathcal{F}^-1[\{F_n\}_{k=0}^{N-1}](k)$ as:
\begin{equation}
f_k =\frac{1}{N} \sum_{k=0}^{N-1} F_n e^{2 \pi i n k/N}
\end{equation}
The DFT reveals periodicities in input data and the relative contributions of the several frequencies in the original function.\\ 
An application of this transform is reported in fig \ref{fig:transform80}. In the left side of the plot a small set of sampled values from a sine wave with a frequency of $80$Hz, in the right its DFT. It has a peak at $80$Hz, a more precise transform can be obtained with a number of points higher ($N$), but clearly the computation of the DFT become more expensive.
It will be often used in the following thesis to show the impact of our models on the signals.\\\\
Another important property of the Fourier transform is that a sampled signal $x(kT)$ is completely determined in the frequency domain by $X(f)$ , where $f \in [0, \frac{1}{T}]$, or equivalently $f \in [-\frac{1}{2T}, \frac{1}{2T}]$.\\
So, let $F_s$ the sampling frequency, it's impossible to distinguish $X(f + nF_s)$ from $X(F_s)$.\\ 
This means that if we don't remove the frequencies higher than the Nyquist frequency, before the sampling process, the Shannon sampling theorem will be violated and an aliasing will be obtained. So the common way to do this operation is put a anti-aliasing filter before the sampling operator, this filter is a \textit{lowpass} filter.
\section{Z-transform}
Another fundamental tool to handle linear time invariant system (LTI) is the \textit{z-transform}. The z-transform of a discrete-time signal $x(n)$ is defined as:
\begin{equation}
 X(z) = \mathcal{Z}(x(n)) = \sum_{-\infty}^\infty x(n) z^{-n}
\end{equation}
where $z$ is a complex variable.\\\\
If $z = e^{i 2\pi f x}$, where $f$ is the frequency and $x$ is the point of evaluation, the z-transform is equal to the discrete Fourier transform.
\[ X(e^{i2\pi f x}) =  \sum_{-\infty}^\infty x(n) e^{-i2\pi f x} = X(f)\]
So, if we consider the discrete Heaviside function:
\[ u(n) = \begin{cases} 0 &\mbox{if } n < 0 \\ 
1 & \mbox{if } n \geq 0 \end{cases}  \]
and we apply the z-tranform:
\[ U(z) = \sum_{n=-\infty}^{+\infty} u(n)z^{-n} = \sum_{n=0}^{+\infty} z^{-n} = \frac{1}{1-z^{-1}} \]
The last passage is correct only if $|z^{-1}|<1$. If we consider $z \rightarrow 1$, $U(z) \rightarrow \infty$. So, we tell that $z=1$ is a \textit{pole} of $U(z)$.\\
The location of the poles of the z-transform are fundamental for the filter design.\\\\
The most important property of the Z-transform, are easily provable considering connection of the Z-transform and the Fourier transform.\\\\
\begin{itemize}
\item \textbf{Linearity}: $x_1(n) + x_2(n) \longleftrightarrow X_1(z) + X_2(z)$, $kx(n) \longleftrightarrow kX(z)$
\item \textbf{Time shift}: $x(n-k) \longleftrightarrow z^{-k}X(z)$
\item \textbf{Multiplication (Convolution)}: $x_1(n)x_2(n) \longleftrightarrow X_1(n)*X_2(n)$
\item \textbf{Convolution (Multiplication)}: $x_1(n)*x_2(n) \longleftrightarrow X_1(n)X_2(n)$
\item \textbf{Differential}: $nx(n) \longleftrightarrow -z\frac{dX(z)}{dz}$
\end{itemize}
The mathematical tools needed for the following arguments have been presented, now we will introduce some notions related to human ear and how it impacts on the acoustic perception.
\section{Human ear and acoustic perception}
The acoustic perception of a sound is a very important branch of the study in the audible system, this because the perception is linked to the ear structure but also to interpretation of this information by the brain. A sound is a continuous analog signal that can theoretically bring an infinite quantity of information.\\
Understandind the features of the audible perception can lead to the design of tools that work very well with our brain.
An excellent example of the importance of these consideration is the codec algorithm  for MP3, this algorithm exploits the weakness and the features of the human ear to delete the inessential information.\\\\
The human auditory system has a complex structure and it performs advanced functions. It is able to elaborate a large number of events but, at the same time, it can be able to identify with precision the pitch and the timbre of the sound or the direction from which it comes.This is possible because lots of functions of the auditory system are performed by the ear but it seems that a large number of elaborations are made by the central nervous system.\\\\
\begin{figure}[h]\centering
\includegraphics[scale=.6]{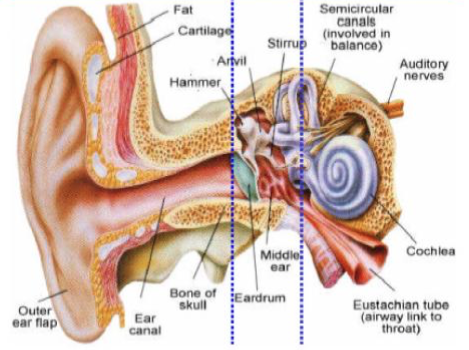} 
\caption{Human ear composition} \label{fig:humanear}w
\end{figure}
The human ear is composed by three parts: external, middle and internal ear, as we can see in fig \ref{fig:humanear}. The external one is formed by the ear flap and the canal. The internal ear is consisted in the hammer, the stirrup and the anvil. These parts are involved in the amplification process.
In the internal ear the main component is the cochlea. It contains the mechanisms to transform the pression variation in the eardrum into nervous impulses, these impulses are interpreted by the brain as sounds.\\
This is possible because the main part of the cochlea is a membrane, called basilar membrane; there are about 30000 nervous receptors distribuited along this part. They convert the movements of the membrane into signals which are sent to acoustic nerve.\\\\
We summarized this notions for a simple reason, when we talk about the human perception, we can't assume that a mathematical representation of the system and the signal can be complete. Banally because there could be some differences between the people, but also because all information are handle by our organs and brain, so when the signal is interpreted by the brain, it could arise some changes respect the original signal.
One of the most important error introduced by the human ear is called \textit{auditory masking}, and an introduction to this behavior can be useful.
\paragraph{Auditory masking}
Our auditory system behaves as a Fourier analyzer, it perceives the individual components of a sound and it distributes them along the basilar membrane in the cochlea. However the vibration peak interests a region that has a dimension, so multiple frequencies can fall in the same region. It causes some imprecision on the perception of the single components.\\
Considering the basilar membrane as a filter bank that works with different frequencies, the discriminatory capacity of our audible system are not absolute, but they depend from the precision of these filters. The phenomenon that allows us to detect the band amplitude of the filters along the basilar membrane is called auditory masking.\\
The masking is the phenomenon that occurs in frequency domain, where a small amplitude signal can become inaudible (masked) by a signal more intense, if these two signals have similar frequencies.\\\\
In general the sounds can be perfectly discriminated from our auditory system only when they fall in two different critical bands, when they fall in the same band, the perception of the differences becomes harder and it is possible only in particular conditions. \cite{soundanalysis}

\subsection{Characteristics of the sound}
If we consider a periodic waveform, the lowest frequency is called \textit{fundamental} frequency (or more simply, fundamental). This frequency gives the biggest contribution in terms of signal information. 
In audible context, each note has a fundamental frequency that characterize it, this is very important for our purpose.\\
An \textit{harmonic} of a periodic waveform is a component of a signal that has a multiple of fundamental frequency as its frequency; so let $f$ the fundamental, it has harmonics  with frequencies $2f$, $3f$, $4f$ and so on.\\\\
In the following the use of the musical terminology will be frequent.\\
We will often refer to the chromatic scale. This is a musical scale with twelve pitches, composed by seven notes and five half tones i.e. the smallest musical interval commonly considered.\\ 
After the twelve pitches a new scale starts. The starting frequency of the next scale is the double of the starting frequency of the previous one. We define an octave as the interval between one pitch and an other with the double of its frequency. The chromatic scale is repeated in ten octaves, so we have 120 different notes in the audible context. The scale is:\\\\
\begin{tabular}{|c|c|c|c|c|c|c|c|c|c|c|c|}
\hline 
Do & Do\# & Re & Re\# & Mi & Fa & Fa\# & Sol & Sol\# & La & La\# & Si\\
\hline 
\end{tabular}\\\\ 
Or, an equivalent writing:\\\\
\begin{tabular}{|c|c|c|c|c|c|c|c|c|c|c|c|}
\hline 
Do & Re\textit{b} & Re & Mi\textit{b} & Mi & Fa & Sol\textit{b} & Sol & La\textit{b} & La & Si\textit{b} & Si\\
\hline 
\end{tabular}\\\\
We can read the symbol with \textit{sharp} for \# and \textit{flat} for \textit{b}.\\\\
We hear a sound when a pressure wave has a particular characteristic, such as intensity or periodicity. The consequence is the perception of a phenomenon composed by three features: \textit{pitch, loudness, timbre}. The interesting thing is that this characteristics is not directly measurable because they are the result of ear and brain elaboration. So, the sound can be interpreted in different way by different people \cite{soundanalysis}.

\begin{figure}[h]\centering
\includegraphics[scale=.6]{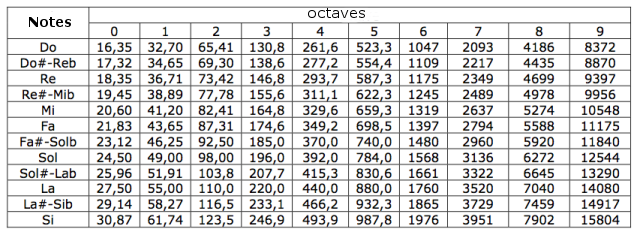} 
\caption{Frequency table of the chromatic scale in Hz.} \label{fig:frequencytable}
\end{figure}

\paragraph{Pitch}
The pitch is the characteristic of the sound that change our perception of it in acute or deep sound. It depends in the largest part by the fundamental frequency, but also by the loudness. The audible range of frequency by the human ear is very high and it is usually considered from 20 Hz to 20 KHz. As we previously explained, it can be some differences between the people in the perception of a sound, some people can hear sound from 16 Hz and their perception can arrive up to 22 KHz; but we will use the range [20 Hz, 20 KHz] as reference.\\\\
Below the lowest threshold, it can be found the infrasound, above the highest threshold the ultrasound. Clearly, each animal species have different range of audible frequency due the different ear characteristics.\\
The lowest note in the chromatic scale has a frequency of $16.35$ Hz, so a very restricted number of people can distinguish this note, the others can hear only the harmonics. The highest note is at the frequency of $15804$ Hz, so no one can hear the harmonics of this note because the first harmonic is at $2f=31608$ Hz, then above the audible range of frequency.  A table can be found in \ref{fig:frequencytable} with all frequency of the chromatic scale. 

\paragraph{Loudness}
The loudness is the sonorous quality associated to the perception of the sound force and it is determinated by the pression that the sound wave exerts on the eardrum. This force is related to the vibration amplitude and to the distance of the point of perception with the sound source.\\\\
To measure the loudness, we will often consider the sonorous level, it can be computed in decibel (dB), as follow:
\[ L_p = 10\log_{10}(\frac{p^2}{p_0^2}) = 20\log_{10}(\frac{p}{p_0}) dB\]
where, $p$ is the mean quadratic pression and $p_0$ is a reference level, called silence threshold. This is the smallest variation of pression that the human ear can distinguish; the standard value is $20\mu Pa$  in the air and $1 \mu Pa$  in the water.\\
The measure with decibel results more appropriate for audible purposes because the human ear response is approximately logaritmic.

\paragraph{Timbre}
The timbre is the quality that, with the same frequency, distinguishes a sound from an other. This feature depends from the sound form and it is related to the superimposition of the foundamental frequency with the various harmonics.\\
In music context, a sound is more complex if it is composed by lots of harmonics, starting from a flute that has a restricted number of harmonics, to arrive at stringed instrument that have a large number of harmonics. In addition to the harmonics other frequencies can be added in the sound, i.e. frequencies that are not integer multiple of the foundamental. If these frequencies are the largest part, we have noise. 

\section{Loudspeakers}
A loudspeaker is a device that convert the electrical audio signal into a sound. In a very simple description, it consists in a set of drivers, each driver is built with a membrane that is handled by a coil of wire. 
The membrane can move freely, within limits, inside of a fixed magnet.\\
The electrical signal passing through the coil create a magnetic field, which produces the mechanical fluctuations of the membrane.\\\\
It is very difficult to create a driver that provides a optimal response with a high frequency range. Due to this problem there are several type of loudspeaker, each driver with particular characteristics has a specific optimal frequency range.
\begin{itemize}
\item \textbf{Sub-woofer and Woofer}: It is dedicated to reproduce the low-pitched audio frequencies (bass), this drivers typically handles the frequency in a range of 20-200 Hz. We used this kind of loudspeaker for the measurements because it creates the most significant error.
\item \textbf{Mid-range}: This kind of drivers reproduces the frequencies in a range of 200-2000 Hz.
\item \textbf{Tweeter}: It typically reproduces frequencies between 2000 Hz and 20000 Hz.
\end{itemize}
\paragraph{Crossover} To divide the input signal into different frequency ranges suited to each driver, there is a set of filters called \textit{crossover}. This is very useful because each driver receive power only for its specific needs, thereby reducing distortion and interference between them. However, as we will see later, it's not possible to realize an ideal filter that rejects all undesired frequencies and allows to pass the desired.\\
There are two types of crossover, passive and active. A passive crossover split the amplified signal. This means that a passive crossover needs no external power but it has some disadvantage, high cost and components with big dimensions.
An active crossover is a filter that divides the signal into individual frequency bands before power amplification, it means that a power amplification for each band will be needed. However, many times, a combination of these techniques is used, for example with a passive crossover for the mid-range and tweeter, and a active crossover for the mid-range and woofer.\\\\
An ideal loudspeaker produce acoustic waves that are a linear transformation of the electrical input signal. This fact is the underlying assumption in traditional analysis of the acoustic system. This kind of approach is based on the frequency response as parameter to evaluate the system and its behavior. So, multiple DFT measurements are used to generate an average system response in the frequency domain. Then, the correction is based on this average response. \\ 
This assumption is based on an oversimplification of the properties of typical dynamic loudspeaker, but such analysis techniques still provide useful insight into speaker performance.\\\\
If we consider the real-world frequency response of a loudspeaker, several contributing factors are linear in nature. Reflections can create either constructive or destructive interference with the direct sound emanating from the speaker cone, leading to peaks and dips in the response. Refraction around the edges of the baffle can create decreased sensitivity at higher frequencies.
Resonances in the speaker assembly and enclosure may create increased sensitivity at discrete lower frequencies. Uncompensated, frequency-dependent impedances in the associated circuitry also lead to uneven sensitivity. All of these factors consequently manifest themselves in the linear transfer characteristic of a loudspeaker; namely, its frequency response \cite{loudspeakerchar}.\\\\
However, the circuits through which the audio signal passes, or the mechanical components responsible for the electro-acoustical transduction may create a nonlinear response.This distortion can be tolerated, or even masked, by the human ear but if they are present in large amounts, they can be unpleasant to hear.\\\\
An interesting and potentially destructive characteristic of the loudspeaker is related to the waves impacting with the membrane. Because these waves are pressure waves, so if they have sufficient force the membrane will be moved. In this sense the loudspeaker behavior is similar to a microphone.\\ 
This is not a trivial observation because a wave bouncing into a wall and returning to the loudspeaker can, potentially creates, some harmonics in the signal due to the membrane movements.\\
This will be relevant for the future developments of this project.

\chapter{Project description}

\section{Motivation of this works}
It is well known that a system which consists of a physical loudspeaker, is a \textit{nonlinear causal system}. In fact, let $x(t)$ be the input of the system, $y(t)$ be the output signal:
\[ y(t) = h(t, x(t)), \]
where $h$ is an nonlinear impulse response added by the system. We will call error the difference introduced by $h(t,x)$.
\[ e(t) = y(t)-x(t),\]
The transfer function $h(t, x)$ contains a large number of distorting factors. In fact an error can be introduced by several phenomena, such as movements of the membrane, the echo produced by the reflection from near objects, other disturbances from several sources.\\
Not all type of loudspeakers are equal in this sense. More specifically, low-frequency drivers produce the most prominent distortions and nonlinearities. \cite{loudspeakerchar} \\\\
Different error types have a different impact on the human ear. The most important forms of distortions in loudspeakers are:
\begin{itemize}
	\item \textbf{Harmonic distortion}: harmonic distortion is made up of harmonic waves in the output which are absent in the original signal.  They can be created by insufficient dynamic range or by nonlinearities in a transfer function.
	\item \textbf{Inter-modulation distortion}: this kind of distortion appears whenever two signals of different frequencies pass through a system with a nonlinear transfer characteristic at the same time. The resulting frequencies are obtained as the sum and difference of the original input frequencies. To the human ear, inter-modulation distortion is the most offensive loudspeaker nonlinearity. Most of these terms are small and can be ignored.\cite{audiosystem}
\end{itemize}

\section{Project aim}
\begin{figure}
\begin{minipage}{.48\textwidth}
\centering
\includegraphics[scale=.33]{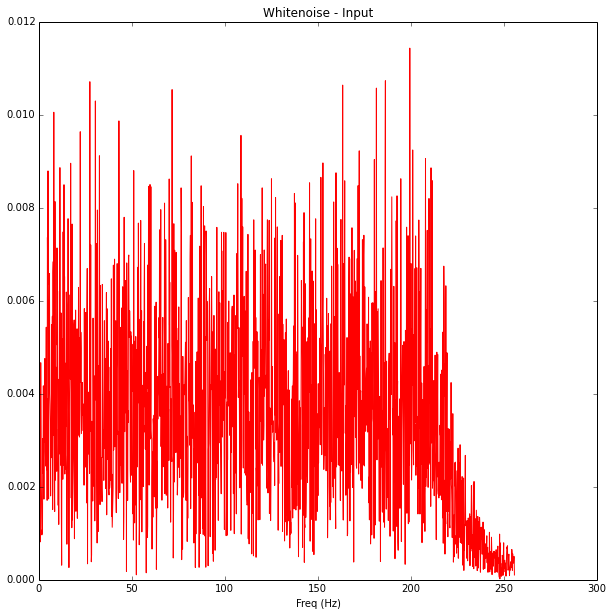} 
\end{minipage}
\hspace{2mm}
\begin{minipage}{.48\textwidth}
\centering
\includegraphics[scale=.33]{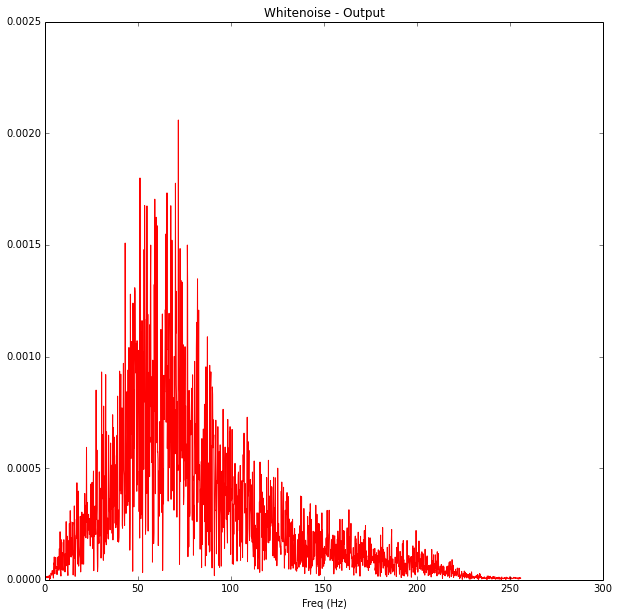} 
\end{minipage}
\caption{Comparison between the input and output spectrum of the loudspeaker used for tests. The spectrum was obtained with a white noise signal.}\label{fig:loudspeaker_behavior}
\end{figure}

The aim of this project is to create a filter that inverts the distorting behavior of the system in order to correct the signal in the pre-processing phase.\\
With this method we want to reduce the harmonics introduced by the system due to the membrane movement.\\\\
Clearly, this kind of inversion must come from the study of a specific loudspeaker, because each speaker has different membrane characteristics, power, and consequently, different harmonic distortions. So, our purpose is to build a mathematical model that encapsulates the loudspeaker features and inverts this model to linearize the output.\\\\
In figure \ref{fig:loudspeaker_behavior} the spectrum of two signals is reported, the input and the output of the loudspeaker. The signal is composed of an equal distribution of frequencies below the 250Hz (white noise). The nonlinear behavior of the system is clear; if our purpose is to obtain a linearization, then the spectrum of the output signal should be similar to the input.\\
\begin{figure}
\begin{minipage}{\textwidth}
\centering
\includegraphics[scale=.5]{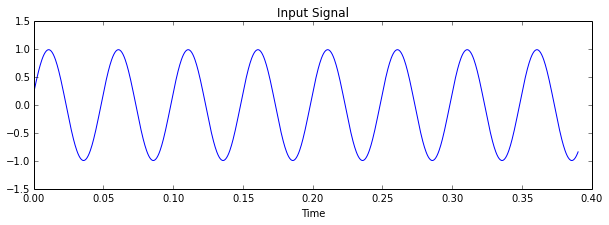} 
\end{minipage}
\hspace{2mm}
\begin{minipage}{\textwidth}
\centering
\includegraphics[scale=.5]{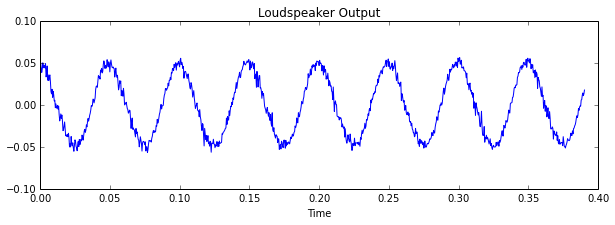} 
\end{minipage}
\caption{Comparison between the input and output of the loudspeaker with a 20Hz monochromatic signal. The nonlinear behavior is clear.}\label{fig:comparison_in_out_20}
\end{figure}

A better idea of the nonlinear disturbance introduced by the system can be observed in fig. \ref{fig:comparison_in_out_20}, where a 20Hz monochromatic signal has been sent to the loudspeaker (first plot), and the output is collected and plotted in the bottom.\\

%TODO AGGIUNGERE LA DESCRIZIONE DELLE PERSONE CHE HANNO LAVORATO AL PROGETTO E IL LABORATORIO CHE HA EFFETTUATO I TEST IN SVIZZERA
%\section{Collaborations}
\section{Implementation Notes}
All prototypes have been implemented with Python using mostly these tools:
\begin{itemize}
\item \textit{iPython} + \textit{SciPy}, to design the filters in first model and collect the results
\item \textit{Audacity}, to analyze the sound
\end{itemize}
To handle the scipy filters we created a Python wrapper to encapsulate these filters into objects. Therefore we exploited abstract classes to handle the filters with different features as the same object.\\
The project has an open source license and is available on GitHub.\\\\
An optimized version of the algorithm has been implemented with \textbf{C++} and \textbf{CUDA} (Compute Unified Device Architecture). CUDA is the parallel computing architecture created by NVIDIA,  and with its extension for C/C++ language, it allows to easily handles the parallelization of code with the GPU (Graphics Processing Unit).
A description of this implementation is presented in this thesis.

\chapter{Digital filters}
Filters are systems through which a signal passes and changes its properties. For example, in the loudspeaker the input electrical signal is converted into a sound pressure wave and the output contains some changes in the information represented by the input, thus the loudspeaker is a filter (unfortunately).\\ 
Other useful filters are graphic equalizers, reverberators, echo devices, phase shifters, and speaker crossover networks.\\\\
A digital filter is a filter that operates on digital signals, it is a function that takes one sequence, that represents the input, and computes a new sequence, the output (or filtered signal). An important property is that a digital filter can do anything that an analog filter can do, the principal difference is that it is implemented by mathematical equations instead of a set of circuits.\\
Clearly, if in the past years we observed a large diffusion of digital filters against the analogical filters there is a reason. Digital filters for their nature have several advantages:
\begin{itemize}
\item a digital filter is programmable, this means the digital filter can easily be changed without affecting the circuitry (hardware).\\
On the contrary, an analog filter can only be changed by redesigning the filter circuit.
\item digital filters are easily designed, tested and implemented. No special hardware is required, all filters presented in this thesis is designed and implemented in a common laptop.
\item the analog filter circuits are subject to electrical current and are dependent on temperature. Digital filters do not have these problems.
\item digital filters are much more versatile than the analog filters, some digital filters have the ability to adapt its behavior to the changing characteristic of the signal. 
\item thanks to the computation capabilities of the actual processors, many complex combinations of filters can be handled in a very simple way. 
\end{itemize}
The simplest types of filters are:\\
\paragraph{Gain}
\[ y(n) = Kx(n) \]
where $K$ is a constant. This simple equation applies a gain factor to each input value. So, if $k>1$ the filters is an amplifier, while $0 \leq K < 1$ the filter is an attenuator. If $k<0$, it corresponds to an inverting amplifier.\\
\paragraph{Delay}
\[ y(n) = x(n-h) \]
where $h$ is the delay. Usually, the first $h$ terms of $y$ is taken equal to zero. 
\paragraph{Two-term difference}
\[ y(n) = x(n) -x(n-1) \]
The output value is equal to the difference between the current input $x(n)$ and the previous input $x(n)$.\\ 
This is interesting because it uses the current input and the past values. It will be seen that this kind of method is necessary to implement complex filters.\cite{introdigital}\\
The \textit{order} of a digital filter is the number of previous inputs used to calculate the current output.
\section{Ideal and Real filters}
\begin{figure}[h]\centering
\includegraphics[scale=.5]{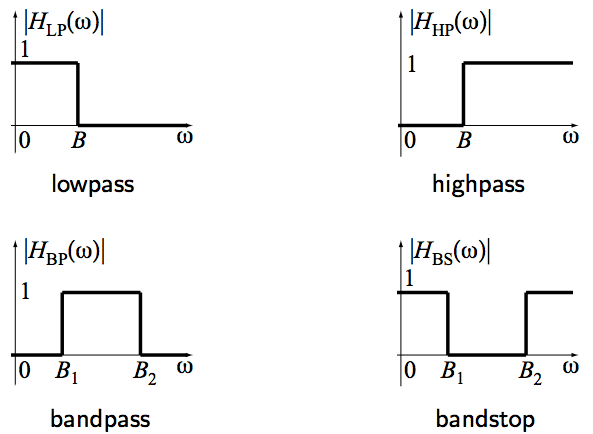} 
\caption{Filters types} \label{fig:ideal_types}
\end{figure}

\begin{figure}[H]\centering
\caption{Comparison between ideal and a real filter}\label{fig:idealreallow}
\includegraphics[scale=.7]{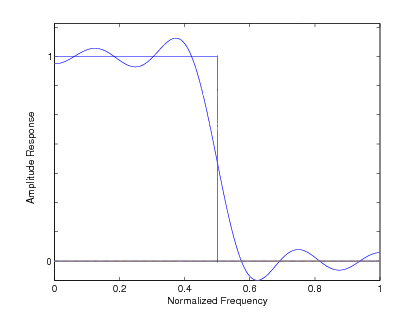} 
\end{figure}

We will look at four kinds of filters, let $f$ the frequency analyzed by the filter:
\begin{itemize}
\item low-pass filters pass all frequencies in the range $|f| \leq B$, for some $B > 0$ and reject all others
\item high-pass filters pass all frequencies in the range $|f| \geq B$, for some $B > 0$ and reject all others
\item band-pass filters pass all frequencies in the range $B_1 \leq |f| \leq B_2$ for some $B_1, B_2 > 0$ with $B_1 < B_2$ and reject all others
\item band-stop filters pass all frequencies in the range $|f| \leq B_1$ and $|f| \geq B_2$ for some $B_1, B_2 > 0$ with $B_1 < B_2$ and reject all others
\end{itemize}
These filter types are illustrated in fig \ref{fig:ideal_types}, however the characteristics must be specified only for a low-pass filter, because the others can be built starting from it.
As defined an \textit{ideal} low-pass filter deletes all frequencies above the \textit{cutoff frequency} while passing those below unchanged.\\\\
The \textit{frequency response} of a system can be defined as the quantitative measure of the output spectrum in response to a input signal; the frequency response of an ideal lowpass filter is a rectangular function as in the figure \ref{fig:idealreallow} (black function).\\
Unfortunately, an ideal filter is unrealizable with a discrete causal system, this because the impulse response of an ideal low-pass filter is a sine wave with infinite extend in time and non-causality. This properties make impossible to implement them in practical application.\\\\
A \textit{real} filter can be realized by truncating the impulse response. The implication is a transition band as a curve and not a vertical line; like in figure \ref{fig:idealreallow} (blue function).
Beside this, a filter need the time to be applied at a signal, the consequence of this fact is a phase shift.\\
Several specifications of these filters have been introduced, they have different transition band, attenuation of Gibbs phenomenon and so on...\\\\
Now, we will analyze a class of filters largely used in practical applications.
\section{Linear filters}
In the linear time-invariant systems (LTI), two relations can be specified for the the operator $\mathcal{H}$ \cite{oppenheim}:
\[ \mathcal{H}(\alpha x_1 + \beta x_2) = \alpha \mathcal{H}(x_1) + \beta \mathcal{H}(x_2) \]
for any $x_1(n), x_2(n)$ and any two scalars $\alpha , \beta \in C$.\\ 
For time-invariant property, we denote the shift operator $\mathcal{T}_r$, it represents the operator that returns the output as input delayed in time of $t$. Then, $\mathcal{H}$ is a time-invariant system if:
\[ \mathcal{T}_r \mathcal{H} = \mathcal{H}  \mathcal{T}_r \forall r \]
Besides, if the linearity property is combined with the representation of signal as delayed impulses as in equation (\ref{eq:convolutionXwithDelta}), reported in previous chapter, it follows that a linear system can be completely characterize by its impulse response $h(n)$\cite{oppenheim}.\\ 
For digital filters, the Z-transform can be used to describe the function, so as the standard way, we will use this form:
\[ H(z) = \mathcal{Z}(h(n)) = \sum_{n=-\infty}^\infty h(n)z^{-n} \]
Considering their impulse response, there are two types of linear filters:
\begin{itemize}
\item \textit{Finite Impulse Response} (FIR): non-recursive filters.
\item \textit{Infinite Impulse Response} (IIR): recursive filters.
\end{itemize}

\begin{figure}[H]
\begin{minipage}{.5\textwidth}
\includegraphics[scale=.5]{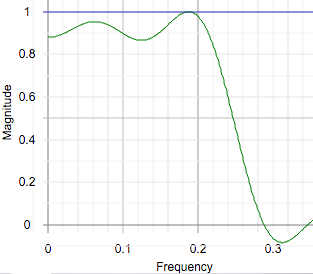} 
\end{minipage}
\begin{minipage}{.5\textwidth}
\includegraphics[scale=.5]{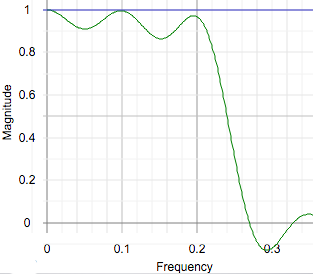} 
\end{minipage}
\caption{Comparison between a FIR filter with order 15 and another one with order 20}\label{fig:fir_example}
\end{figure}

\begin{figure}[H]\centering
\includegraphics[scale=.6]{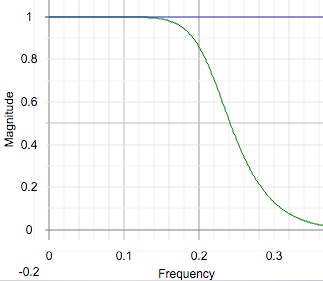} 
\caption{IIR order 4} \label{fig:iir_example}
\end{figure}

\paragraph{Finite Impulse Response} 
We can define the equation of a FIR filter as a weighted sum of the last $N+1$ input values, where $N$ is the order chosen for the filter.
\begin{equation}
y(n) = b_0x(n) + b_1x(n-1) + ... + b_Nx(n-N) = \sum_{i=0}^N b_ix(n-i)
\end{equation}
It can be noted that the application of a filter is a discrete convolution between the input vector and the vector of coefficients.\\
The impulse response of this kind of filter is:
\begin{equation}
h(n) = \sum_{i=0}^N b_i \delta(n-i) = \begin{cases} b_n &\mbox{} 0\leq n \leq N \\ 
0 & \mbox{} otherwise \end{cases}
\end{equation} 
In these filters, the impulse response has a finite duration, this means that it is equal to zero in a finite time. In fact, from the previous equation, it becomes equal to zero in $N$ steps.\\
The z-transform of the filter is:
\begin{equation}
H(z) = \sum_{i=0}^N b_i z^{-i}
\end{equation}
FIR filters have several properties:
\begin{itemize}
\item no feedback is required, this means that only input signal is used during the application of the filter. As we will see, this is the difference between FIR and IIR filters. However, this application allows to avoid the forwarding of the rounding error.
\item are inherently stable, since the output is a sum of a finite number of finite multiples of the input values.
\item if we use a coefficient sequence symmetric, the phase of the signal will change only linearly. This properties is very important in sound application, because a linear phase is simpler to handle and the risk to occur in a phase cancellation problem is smaller.\\
For this reason all linear filters used in this thesis are FIR filters.
\end{itemize}
However, FIR filters have also some disadvantage compared to IIR filters, the principal one is the computational cost; to obtain a behavior similar to IIR filter in the frequency cut, higher orders of the filter is required.\\\\
\paragraph{FIR filter example}
Here, a simple example of low-pass FIR filter with its frequency response, the changes in the precision of the filter can be seen in fig \ref{fig:fir_example}, if the order become higher.
\paragraph{Infinite Impulse Response} In this type of filters the response to an impulse does not go to zero after a certain point, but can grow indefinitely. This happen because IIR filters use in addition to the input signal (\textit{feedforward components}), even a part of past output signal (feedback components). If we consider the difference equation of these filters, they can be defined as:
\[ y(n) = \frac{1}{a_0} (b_0x(n) + b_1x(n-1) + ... + b_Ix(n-I) \]
\[- a_1y(n-1) - a_2y(n-2) - ... - a_Oy(n-O))\]
where, $I$ is the feedforward filter order, $O$ is the feedback filter order.  $b_i$ and $a_i$ are the coefficients of the filter respectively for the feedforward and feedback part. Often in literature it is expressed with the condensed form:
\begin{equation}
y(n) = \frac{1}{a_0} (\sum_{i=0}^I b_ix(n-i) - \sum_{j=1}^O a_jy(n-j))
\label{eq:iirdefinitioneq}
\end{equation}
we can rewrite eq. \ref{eq:iirdefinitioneq} moving the output part in the left of the equation:
\[ \sum_{j=0}^O a_jy(n-j) = \sum_{i=0}^I b_ix(n-i) \]
and the z-transform of the equation (\ref{eq:iirdefinitioneq}) is:
\[ \sum_{j=0}^O a_jz^{-j} Y(z) = \sum_{i=0}^I b_iz^{-i}X(z) \]
The transfer function of the filter is:
\[ H(z) = \frac{Y(z)}{X(z)} = \frac{\sum_{i=0}^I b_iz^{-i}}{\sum_{j=0}^O a_jz^{-j}} \]
In general the coefficient $a_0$ is taken equal to $1$, and the final form of the transfer function become:
\begin{equation}
H(z) = \frac{\sum_{i=0}^I b_iz^{-i}}{1 + \sum_{j=1}^O a_jz^{-j}}
\end{equation}
This type of filters have advantages in term of computational resources required for the filter applications. It is possible to obtain a good filters specifying constraints for the pass-band, stop-band and ripple. The order requirements are usually lower than the FIR filters, this difference is often large.\\
However, IIR filters have a nonlinear phase response it could suffers of instability problems due to possible zero in the denominator of $H(z)$.
\paragraph{IIR filter example}
\textbf{Order = 15} Now, we present an example of low-pass IIR filter of 4-th order, we obtain a relative good precision compared to the previous FIR filter presented, fig \ref{fig:iir_example}.
There are several method, mainly based on the \textit{windowing}, to design the digital filters, both FIR and IIR. The idea is create a window of the signal, so take only one part of the signal, and solve the system for our conditions. With a iterative process, make the window smaller to achieve a better precision.

\section{Multirate Signal Processing}
Multirate signal processing is a technique employed to handle signals with different sampling frequency. It comes to help whenever conversions among different frequencies are needed, it is also a signal processing tool in its own, with many useful applications in the design of efficient filtering schemes.\\
In the following this method will be used in both presented ways, the so called brute-force model is based on a multirate filter bank, i.e. a set of filters that work with different frequencies. \\
In the second approach with the Volterra series, multirate signal processing is used to decrease the computational cost of the nonlinear approach\\
\subsection{Decimation}
The downsampling process, also called \textit{decimation}, is a method to reduce the sampling frequency of a signal. In particular, a constant $N$ has to be choosen, this constant is called \textit{decimation rate} and we define an operator to handle the decimation process, called \textit{decimator}, defined as:
\[ \mathcal{D}_N(x(n)) = x(nN) \]
Downsampling discards $N-1$ out of $N$ samples, a particular attention should be put on the choice of the decimator factor because it could imply the introduction of signal distortion.
\begin{figure}[H]\centering
\includegraphics[scale=.6]{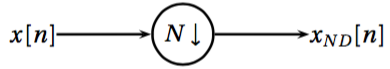} 
\end{figure}
\paragraph{Properties of $D$}
Consider a signal composed by the numbers:
\[ x(n) = ..., x(-2), x(-1), x(0), x(1), x(2), ... \]
If we consider the decimator with $N=2$ we have:
\[ x_{ND}(n) = \mathcal{D}_2(x(n)) = ..., x(-4), x(-2), x(0), x(2), x(4), ... \]
now, if we consider a shifted time version,we have:
\[ \mathcal{D}_2(x(n+1)) = ..., x(-5), x(-3), x(1), x(3), x(5), ... \]
For the definition,  $\mathcal{D}_2(x(n+1)) = x(2n+1)$.\\
this means that the downsampling operator is not time-invariant, in particular the decimator is a \textit{periodically time-varying} operator. It is possible to show the the downsample operator is linear.\cite{signalcommunications}.

\begin{figure}[h]\centering
\includegraphics[scale=.4]{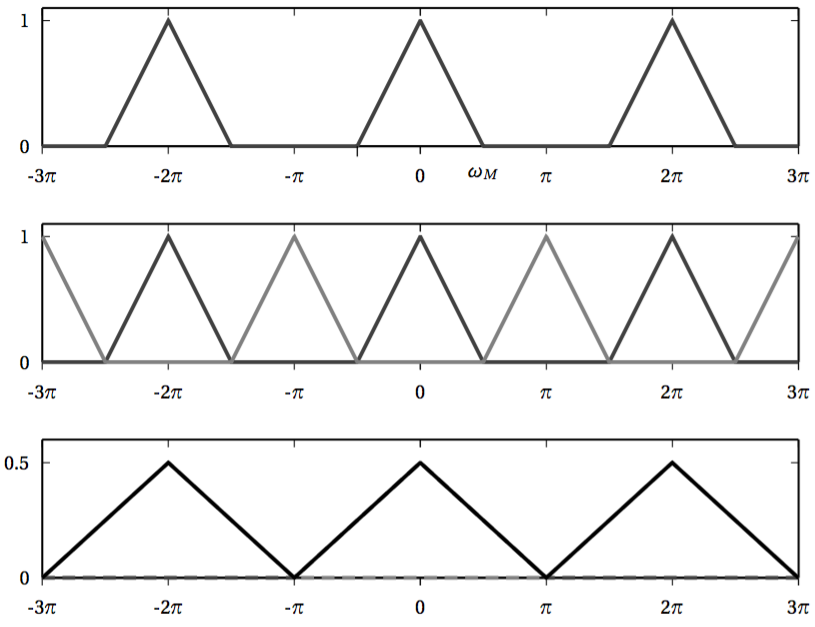} 
\caption{Example 1: In the top picture $X(n)$, i.e. the spectrum of original signal. In the second figure the signal given by the summation parts. In the last picture the final signal obtained as results of the summation. In this case, a good decimation is obtained.} \label{fig:example1}
\end{figure}

\begin{figure}[h]\centering
\includegraphics[scale=.4]{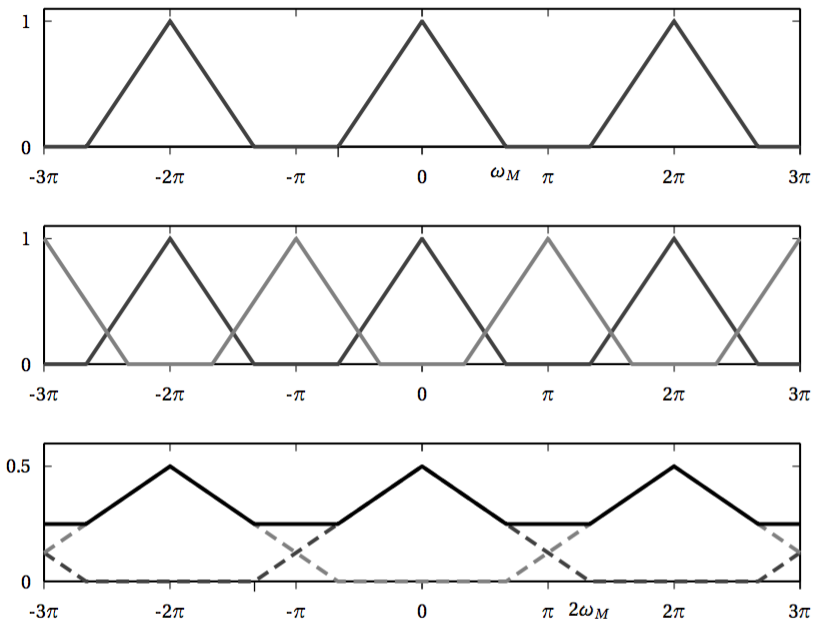} 
\caption{Example 2: In this second figure we can see that the summation parts in this case overlap. The consequence is the creation of wrong final signal (last picture).} \label{fig:example2}
\end{figure}

\paragraph{Constraints} Consider the decimated signal $ x_{ND}(n) = x(nN)$, $-\infty < n < +\infty$, and the sampling function $\delta_N$ defined as:
\begin{equation}
\delta_N(t) = \sum_{-\infty}^{\infty} \delta (t-nN)
\label{eq:samplingfunction}
\end{equation}
Where $\delta (t)$ is the \textit{Dirac} function defined in equation (\ref{eq:deltadirac}).\\
$x_{ND}$ can be specified exploiting the equation (\ref{eq:samplingfunction})
\begin{equation}
x_{ND}(n) = x(n)*\delta_N(n)
\end{equation}
A convolution in the time domain implies a multiplication in the frequency domain, then:
\begin{equation}
X_{ND}(f) = \frac{1}{N} \sum_{k=-\infty}^{\infty} X(f-\frac{k}{N}) = f_s \sum_{k=-\infty}^{\infty} X(f-k f_s)
\label{eq:samplingfrequency}	
\end{equation}
Where $f_s$ is the sampling frequency related to $N$.\\\\
Equation (\ref{eq:samplingfrequency}) implies that the resulting spectrum is the scaled sum of $N$ superimposed copies of the original spectrum; each copy is shifted in frequency by $k f_s$ and stretched by $f_s$.\\  
For a band-limited signal with band $B$, the \textbf{Nyquist} condition can be defined:
\begin{equation}
\frac{1}{f_s} = N \leq \frac{1}{2B}
\end{equation}
No aliasing occurs if the \textbf{Nyquist} condition is satisfied, otherwise the spectral copies of the original spectrum overlap and the reconstruction of the signal causes aliasing \cite{signalcommunications}.\\\\
Two examples of decimation are presented with the help of images: the first part of the figures shows the initial spectrum of $X(f)$; the second part shows, in different shades of gray,  the individual components of the sum in the last equation, before scaling and stretching; the third part present the final signal.\\
\paragraph{Example 1} Consider a decimate factor $N=2$; if we consider a signal with the maximum frequency $\omega_M = \pi /2$, the non-aliasing condition is fulfilled and the shifted versions of the spectrum do not overlap, fig \ref{fig:example1}.
\paragraph{Example 2} Now consider an other signal with the maximum frequency  $\omega_M = 2\pi /3 > \pi/2$, in this case the Shannon condition is violated and the spectral copies do overlap as in fig. \ref{fig:example2}.
\paragraph{Anti-aliasing} To avoid the overlapping problem, an additional filter is inserted before the decimator, an \textit{anti-aliasing filter}. This is a low-pass filter with cut-off frequency $\omega=\pi/N$, in this way the parts of the signal frequencies that could overlap are removed.
\begin{figure}[H]\centering
\includegraphics[scale=.6]{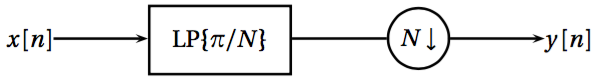} 
\end{figure}

\subsection{Upsampling}
\begin{figure}[h]\centering
\includegraphics[scale=.5]{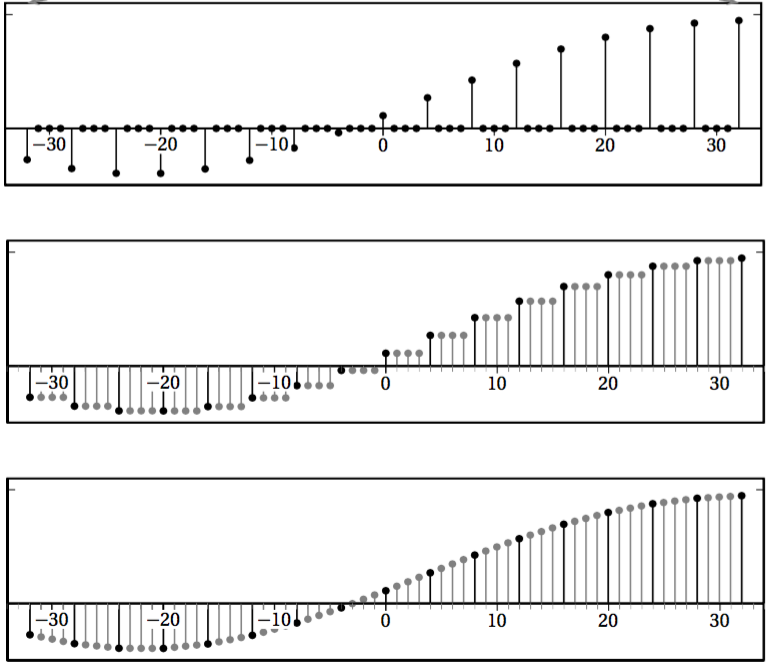} 
\caption{Upsampling with interpolation process, using a low-pass filter. In the first image the zero-interpolation, in the second image the first-order interpolation and in the last the first-order interpolation with low-pass filter after the upsampling.} \label{fig:interpolationlowpass}
\end{figure}

The simplest operation to perform an upsampling consists in inserting $N-1$ zeros between every two input samples; so if we call $\mathcal{U}$ the upsampling operator, we have:
\[ x_{NU}(n) = \mathcal{U}_N(x(n)) = \begin{cases} x(k) &\mbox{for }  n=kN, k \in Z \\ 
0 & \mbox{otherwise } \end{cases} \]
The spectral description of upsampling is expressed with the z-transform domain:
\[ X_{NU}(z) = \sum_{n=-\infty}^\infty x_{NU}(n)z^{-n} \]
\[ = \sum_{k=-\infty}^\infty x(k)z^{-kN} = X(z^N)\]
And the Fourier transform is:
\[ X_{NU}(e^{j\omega}) = X(e^{j\omega N}) \]
the upsampling is a contraction of the frequency axis by a factor of $N$.\\\\
With upsampling defined as the previous equation, the signal suffers from a drawback. In time domain, there are $N-1$ zeros between every sample drawn from the input. An interpolation process can be used to avoid this behavior, a lowpass filter with cut-off frequency $\pi/N$ can be used, as shown in \cite{signalcommunications} and in fig \ref{fig:interpolationlowpass}. \\\\ 
A fractional decimation rate can be achieved combining the downsample and upsample. To avoid aliasing an upsampler must be put in cascade with the lowpass filters and the downsampler.
\begin{figure}[H]\centering
\includegraphics[scale=.4]{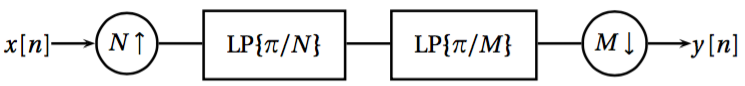} 
\end{figure}
\section{Adaptive and Nonlinear filters}
In the previous section we have discussed about linear and time-invariant filters (LTI). Now we will focus on the time-invariant hypothesis, this means that the filter coefficients remain fixed, for the entire operation of the filter.\\
This characteristic is not always a good property, in our case, a loudspeaker is subjected to an evolution of its properties. The membrane changes its elasticity, the box can change its resonances. The echo from the room changes frequently and so on.\\
An \textit{adaptive} filter has coefficients which are designed to vary in the time, this adaptive process has to be made in a strictly controlled manner to avoid dangerous cases.\\\\
Usually, the variation of the coefficients is low. It is like to have a filter with a good approximation of the system behavior and able to change its properties in during the filters life, and its evolution is made by small steps.\\
To build this kind of filter, we have to monitor the input and output signal, calculate the filter error and correct it.\\\\
There are several methods which allow to reduce the error between the desired and output signals, assuming to have a suitable reference signal. The most difficult decision is what algorithm should be used; this is an optimization algorithm that try to find the minimum of the error function.\\\\
The most widely used adaptive algorithm is the Least Mean Square (LMS) algorithm, it easy to implement and understandable. This algorithm will be used in the following with a modification related to the step size.\\\\
In addition, a nonlinear function can be used for the filter, and in some cases this is fundamental because the linear assumption is far from the reality.\\
However, the nonlinear filters are more complex and the computation is heavier.
\subsection{Active Noise Control}
\begin{figure}[h]\centering
\includegraphics[scale=.4]{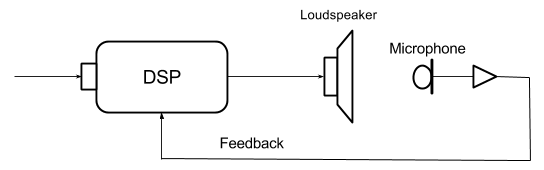} 
\caption{Feedback control}\label{fig:feedback}
\end{figure}

A very useful method is called \textit{Active Noise Control} (ANC), and it aims to delete the noise coming from external sources. ANC has wide application to problems in manufacturing, industrial operations, and consumer products.\\
The most used architecture for these applications is a mixing of adaptive linear filters and feedback control. A feedback channel utilises of a microphone in front of loudspeaker to measure the output, this is sent to the loudspeaker as additional information (feedback channel), as in fig. \ref{fig:feedback} .\\\\
Active noise control exploits the principle of superposition of anti-noise of equal amplitude and opposite phase on primary noise, thus resulting in the cancellation of both signals.
The feedback signal, that contains original signal, harmonics and noise created by the environment, is processed by a filter to extract the noise. This error signal is subtracted in anti-phase from the next input.\\\\
Adaptive filters are developed considering that a noise is not fixed, but these methods are very simple to apply in a controlled environment, and during the laboratory tests, their results are very reliable. It is more difficult when applied in real situations, also because it's not always easy to determine "a priori" what is the noise.\\\\
This is relatively simple when the noise is a constant sound, like a train in movement or a constant background speaking by people. The average noise sound can be removed safely; but if we consider for example a clacson sound, this is an impulse and it isn't possible to remove it, but if one try to remove this noise from the next input, probably a destructive effect is obtained.\\
A good review of these algorithms can be found in \cite{reviewANC}.
\part{Design and implementation}
\chapter{Error correction: Piecewise linear approach}
\begin{figure}[h]\centering
\includegraphics[scale=.35]{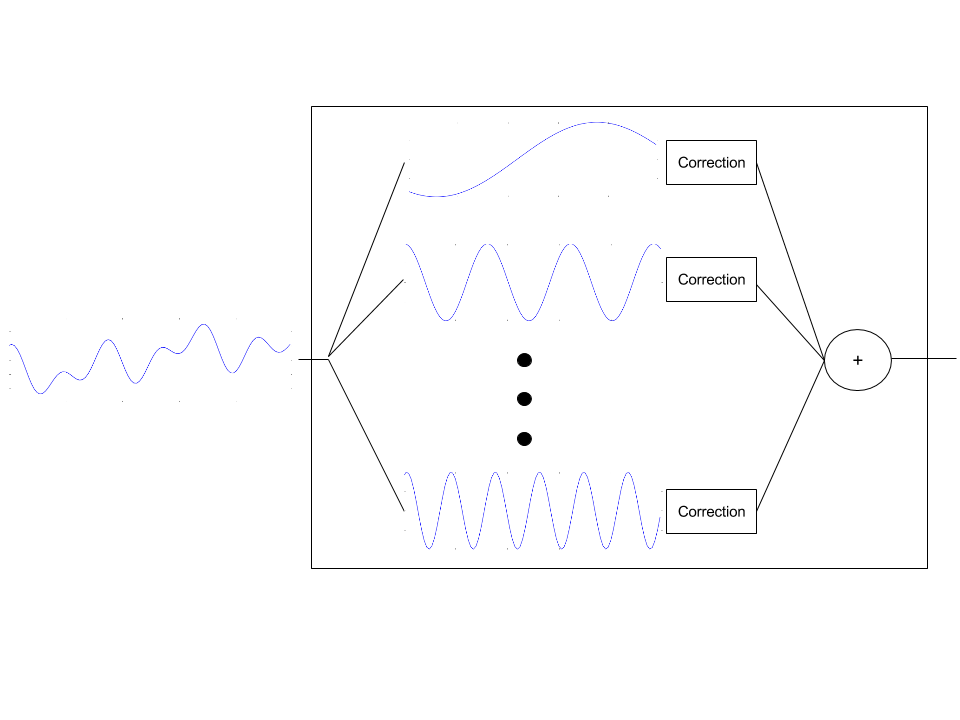} 
\caption{Scheme of a DSP that works with brute force approach} \label{fig:bruteforcemodel}
\end{figure}

In this chapter, it will be described the first approach used to solve the problem of the loudspeaker signal correction. In this method the nonlinear operator is broken up in several parts, one for each note. Within the each part, we assumed a linear behavior.\\
This approach is widely used in methods for nonlinear systems, it is called \textit{linearization}. A set of system points are identified and near these points, if several conditions are valid, linearity can be assumed.\\\\
The logic of linearization method has been applied to the sound, an audio signal contains a set of frequencies that represent a set of notes.\\
An algorithm will be presented to split the signal in a set of frequency regions in which a linear filter can be used, in order to correct the error.\\
Using a measurement system to collect the loudspeaker output, the error related to each note can be computed.\\
\paragraph{Assumption:} As shown in \cite{loudspeakerlinearization}, the loudspeaker can be modeled as a weakly nonlinear system. In these systems the first order dominates and there are some orders of difference with the contributions of the nonlinear terms. Then, let $f_x$ the frequency considered, it can be made the assumption of local linearity if the interval $[f_x-\delta, f_x + \delta]$, $(\delta>0)$  is sufficiently small.\\\\  
Let $f_x$ the frequency of the note $x$, the error related to this note is denoted as $e_x$.\\
The algorithm must work as follow:
\begin{enumerate}
\item split the sound into a set of frequencies, one for each note.
\item apply the correction signals, based on $e_x$
\item sum all corrected signals and return the output.
\end{enumerate}
The scheme of figure \ref{fig:bruteforcemodel} is a graphical representation of this algorithm.\\
We used a \textbf{filter bank} to separate the sounds in a set of sub-signals. This structure handles a set of filters as a single object.
\section{Filter Bank Design}
Filters should extract the component expressed in frequency for each note; a set of \textit{pass-band} filters have been designed for this purpose. Considering 12 notes (tones + semitones) and 10 audible octaves, filter bank must contain 120 filters.\\
To design the filters, we consider the entire range of audible frequencies, with sampling frequency of 44100 Hz. \\\\
Each filter is different from the others, due to the cut off frequency and the position related to the zero frequency. However, common features have been imposed for the filters, this is necessary in the following to handle signals in the same way.\\
We fixed two parameters with the same values, regarding the behavior and the attenuation in the pass band and the stop band. The attenuation is measured in $dB$, the reference value is the amplitude of the pass band.
\begin{itemize}
	\item Passband ripple, denoted with $rp$. This is the fluctuations, or variations, in the frequency magnitude response within the passband of a filter; fixed to $-3$ dB.
	\item Stopband ripple, denoted with $rs$: This is the minimum attenuation required in the stop band, fixed to $-40$ dB.
\end{itemize}
These values have become a standard in practical audio applications \cite{dsppratical}.\\\\
We used FIR filters in order to benefit of the linear phase response, with this feature is easier avoid the phase cancellation problem. Besides, these filters are always stable, unlike for IIR filters.\\
Unfortunately, FIR filters are computationally more expensive, we need to have a higher order filter to obtain the same precision on cutting frequencies.
\begin{figure}[H]\centering
\includegraphics[scale=.6]{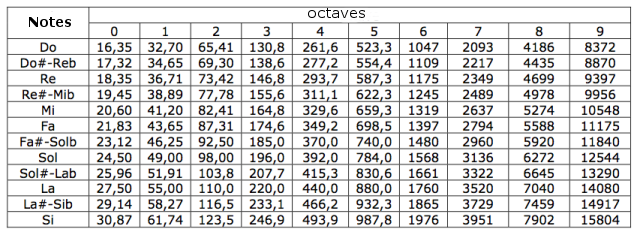} 
\caption{Frequency table of the chromatic scale in Hz.} \label{fig:frequencytable2}
\end{figure}
Fig. \ref{fig:frequencytable2} reports the distance in frequency between a note and the next. In the first octaves, it is extremely small (1-5Hz) if compared with the sampling frequency (44100Hz); this has caused a problem during the design of the filters.\\
In the next paragraphs we will use again the term \textit{small} and \textit{near} concerning the frequency space, this must be interpreted as if compared to the sampling frequency. Besides, we want to maintain acceptable the computational cost of the filter bank; then we will use filters with order less than $100$.

\begin{figure}[h]
\centering
\includegraphics[scale=.5]{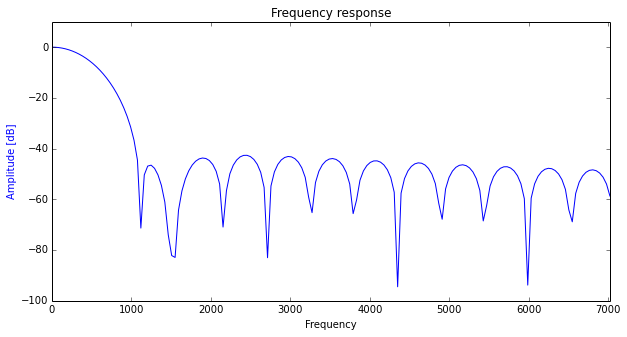} 
\caption{Frequency response of a passband filter designed for interval [$22.19Hz$,  $24.04Hz$] and sampling frequency $44100 Hz$. The passband filter is degenerated to a unprecise lowpass filter due to the high difference in frequency between filter frequencies and sampling frequency.} \label{fig:passbanddegenered}
\end{figure}

\paragraph{Problem} 
If the pass band is small and the frequency is near to zero; it's not possible to design a passband filter, with acceptable order, for our purpose. As it can be seen in figure \ref{fig:passbanddegenered}, the transfer function can not remove the frequency less than $f_x$ and fall again in the stopband in a small frequency space. This happens because a real filter hasn't transition curve vertical.\\
The consequence is a degeneration of the passband filter in a lowpass filter. To solve the problem, we adopted multirate filtering.
\subsection{Multirate Filter Bank}
\begin{figure}[h]
\begin{minipage}{\textwidth}
\centering
\includegraphics[scale=.4]{img/WRONG_LINEAR_FILTER.png} 
\end{minipage}
\begin{minipage}{\textwidth}
\centering
\includegraphics[scale=.4]{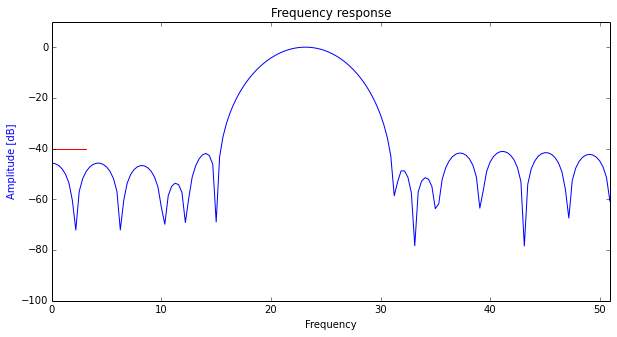} 
\end{minipage}
\caption{Filter design improvements with multirate approach, example with $[22.19Hz$,  $24.04Hz]$ - downsampling from $44100Hz$ to $320 Hz$.} \label{fig:multirate_result}
\end{figure}

The previous problem is linked to the sampling frequency, too high compared with the frequencies handled by the filters. Using decimation, the signal can be scaled in a frequency range more appropriate.\\
Considering again the frequency table in figure \ref{fig:frequencytable2}, the highest frequency related to the second octave is $f_a=61,74Hz$, the Shannon sampling theorem asserts that: to preserve the information without aliasing, it's sufficient to use a sampling frequency greater than $2 f_a$.\\
It's useless have a sampling frequency of $44100Hz$ for these filters.\\\\
A set of frequency regions has been identified, called bands. They separate the spectrum in a set of intervals $[B_1,B_2, ..., B_i]$. Within the same band, a common sampling frequency is used. It's greater than the double of the maximum frequency $B_i$, in order to preserve the constraint of the Shannon sampling theorem.\\ 
The filtering algorithm becomes:\\\\
Let $f_{B_i}$ the sampling frequency of the band $B_i$:
\begin{enumerate}
	\item $\forall B_i,$ execute the downsampling of input to $f_{B_i}$
	\item split the sound contained in $B_i$ into a set of frequencies, one for each note.
	\item apply the correction signals, based on $e_x$
	\item execute the upsampling for all signal to a common frequency
	\item sum all corrected signals and return the output.
\end{enumerate}
We report in the table below the frequencies used:\\
\begin{tabular}{|c|c|c|c|}
\hline 
Min. frequency & Max. frequency & Octaves & Sampling Frequency \\ 
\hline 
0 Hz & 61.64 Hz & 1-2 & 300 Hz \\ 
\hline 
65.41 Hz & 249.9 Hz & 3-4 & 700 Hz \\ 
\hline 
261.6 Hz & 987.8 Hz & 5-6 & 2940 Hz \\ 
\hline 
1047 Hz & 3951 Hz & 7-8 & 11025 Hz \\ 
\hline 
4186 Hz & 15804 Hz & 9-10 & 44100 Hz \\ 
\hline 
\end{tabular} 
\vspace{10mm}
\\In fig \ref{fig:multirate_result}, the comparison between the wrong passband filter with cut off frequencies $\in [22.19Hz$,  $24.04Hz]$ and a sampling frequency of $44100Hz$, and the new filter with the same cut off frequencies but a sampling frequency of $320Hz$.

\paragraph{Precision:}
To understand the best configuration, several characteristics have been tested during the design of the filters.\\
A further consideration is related to the distance between a note and the next; this space is not fixed. For example, in the frequency table can be seen the distance between the first note and the next semitone. It is $0.97$Hz in the first octave, but in the last one, it is $498$Hz.\\
This feature can be exploited for our purposes, the spectrum has been divided in frequency intervals and filters with different complexity have been used. This method allows to obtain narrower filters for lower frequencies and progressively larger with higher frequencies.\\\\
Below, the common specifications of the filters have been presented, considering the frequency range in which the note belongs. The order is reported and the frequency interval of the transition band; this specification ensures that outside this interval, the signal is attenuated below $-40$db: this limit is taken to consider an audio signal negligible\\
Let $f_x$ the frequency to extract:
\begin{itemize}
	\item $f_x \in [0Hz, 500 Hz]$ - order 50 - transition in $[f_x - 3, f_x + 3] $
	\item $f_x \in [500Hz, 2000Hz]$ - order 30 - transition in $[f_x - 5 , f_x + 5 ] $
	\item $f_x \in [2 KHz, 20 KHz]$ - order 20 - transition in $[f_x - 10 , f_x + 10 ] $
\end{itemize}
The range of transition band is approximated.
\subsection{Test}
\begin{figure}[h]
\centering
\begin{minipage}{\textwidth}
\includegraphics[scale=.6]{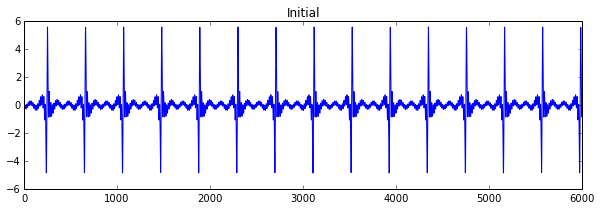} 
\end{minipage}
\hfill
\begin{minipage}{\textwidth}
\includegraphics[scale=.6]{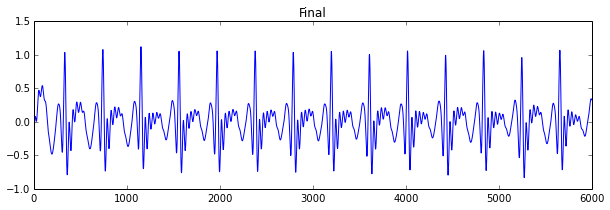} 
\end{minipage}
\caption{Comparison between the initial and final signal, obtained with a linear filter bank. The result is very different from the expected result, a pretty similar signal should be observed.}\label{fig:badFilterResult}
\end{figure}

The test signal is chosen in according to the loudspeaker used for test, the same signal will be used in the following to test the loudspeaker model. A subwoofer has been used, then we created a test signal with low frequencies.\\\\
To understand the goodness of this approach, a particular test has been done. A signal obtained as superimposition of sine waves, with frequencies less than $200$Hz, has been divided in the frequency bands with the filter bank. For this step, we used the downsampling specifications presented previously. Therefore these sub-signals have been filtered with an upsampler and summed together, without any correction.\\
The expected result of this test is a signal very similar to the initial signal, with the amplitude scaled due to the filter attenuation coefficient.\\\\ 
In fig. \ref{fig:badFilterResult}, it can be seen the comparison between the initial and final signal.\\
Looking the results, it can be verified that this method turns out to be very inaccurate. The transformation doesn't consist only in a linear attenuation.\\
The imprecision is caused by the non vertical transition curve of real filters. This implies that in addition to the interested note, others frequencies are taken by the filters. Two consequences derive from this problem:
\begin{itemize}
\item It's impossible to correct a single note, the change is done also in other frequencies; the linear assumption about the filters becomes difficult to satisfy.
\item If the transition band of two adjacent filters operates on a common frequency space, the signal related to these frequencies is added two times in the final signal. This implies a constructive and destructive effects, producing an unwanted behavior.
\end{itemize}
Looking at the distance between the notes, a solution could be to enlarge the working region of the filters, considering some notes together.
This would reduce surely the second problem, reducing the overlapping frequency regions. In this case the interval $\delta$ of the frequency would become larger, and the linearity assumption breaks down.\\\\
The method developed up to now has been very useful to understand the features of the problem. At this point, the approach must be changed to achieve a more effective filter.\\\\
The idea of the new method is based on a different reasoning. Instead of considering directly the sound produced as output, the loudspeaker is analyzed as a physical object. In fact, the error contained in the output signal is strictly dependent from the loudspeaker properties. If its features can be described by a model, its behavior can be inverted.\\
This new model is based on the \textbf{Volterra series} that will be presented in the next chapter.

\chapter{Volterra Model}
\section{Volterra series}
In this chapter, it will be presented the final approach used in the loudspeaker signal correction. The idea is the creation of a mathematical model of the system, it will be used to invert the loudspeaker behavior.\\\\
The tool used to represent the system is a series, called Volterra series. This model has been widely studied in theoretical way as method to handle weakly nonlinear system.\\
We chose the Volterra series because several researches proved its potentiality in control systems, with a special attention to the loudspeakers.\\\\
Let $H$ a nonlinear continuous time-invariant system, with an input signal $x(t)$ and an output signal $y(t)$, defined by the equation:
\begin{equation}
y(t) = H(x(t)) = y_0(t) + y_1(t) + y_2(t) + y_3(t) +...
\label{eq:general_nonlinear}
\end{equation}
Where $y_i$ is the contribution to $y$ related to the $i$-th terms of the series.\\
Then, $H$ can be exactly expressed with the Volterra series as \cite{volterra} :
\begin{equation}
y(t) = h_0 + \sum_{n=1}^{+\infty} \int_a^b ... \int_a^b \! h_n(\tau_1, ..., \tau_n) \prod_{j=1}^n x(t-\tau_j) d\tau_j
\label{eq:volterra_definition}
\end{equation}
Where $h_n(\tau_1, ..., \tau_n)\neq 0$, it's called $n$-th order kernel of Volterra. The kernel can be considered as a generalization of the impulse response used in linear system.\\
The equation (\ref{eq:volterra_definition}) can be rewritten using (\ref{eq:general_nonlinear}) as:
\[ y_0 = h_0 \]
\[ y_1 = \int_a^b h(\tau_1) x(t-\tau_1 )d\tau_1 \]
\[ y_2 = \int_a^b \int_a^b  h(\tau_1, \tau_2) x(t-\tau_1 ) x(t-\tau_2 )d\tau_1 d\tau_2 \]
\[ ...\]
A symmetric kernel is defined by the relation:
\begin{equation}
h_{sym}(\tau_1, ..., \tau_n) = \frac{1}{n!} \sum_{\pi} h_n(\tau_{\pi (1)}, ..., \tau_{\pi (n)} )
\label{eq:symmetric_kernel}
\end{equation} 
Where  $\pi $ denotes any permutation of the integers $1, . . . ,n$ and the indicated summation is over all $n !$ permutations of the integers 1 through n.\\
As shown in \cite{volterra}, without loss of generality the kernel of a Volterra system can be assumed to be symmetric.\\\\ 
The Volterra expansion represents a generalization of the Taylor series which can be used when the output is dependent only from the input at current time.\\
The main feature of the Volterra series is the ability to capture the \textit{memory effect} of a system; in these situations, the output value can be expressed as linked to all past values of the input.\\\\
Considering the equation (\ref{eq:volterra_definition}), a special attention should be put on the parameters $a, b$ that represent the memory of the filters. As we will see, the memory will impact on the precision and the performance of the filter.\\
Let for example, a system $T$ with input $x(t)$, $T$ returns an output $y(t)$ dependent from the past $100$ seconds.\\ 
If we build a model for this system with a memory of $5$ seconds, the model output will be probably an approximation far from the real value.\\
If a memory of $90$ seconds is used, we will obtain in general a more reliable value. \\\\
For any application of this method, we need to consider the truncated series. Then, let $P$ the maximum order considered for the system, it must be used:
\begin{equation}
y(t) = h_0 + \sum_{n=1}^{P} \int_a^b ... \int_a^b \! h_n(\tau_1, ..., \tau_n) \prod_{j=1}^n x(t-\tau_j) d\tau_j
\label{eq:truncated}
\end{equation}
The equation (\ref{eq:truncated}) can be used in a discrete-time domain, replacing the integrals with summations:
\begin{equation}
y(n) = h_0 + \sum_{p=1}^P \sum_{\tau_1=a}^b ... \sum_{\tau_p=a}^b \! h_p(\tau_1, ..., \tau_p) \prod_{j=1}^p x(n-\tau_j)
\end{equation}
As shown in \cite{loudspeakerchar}, exploiting the symmetric kernel definition in (\ref{eq:symmetric_kernel}), the Volterra expansion can be rewritten as:
\[
y(n) = h_0 + \sum_{\tau_1=0}^{M-1} h_1(\tau_1)x(n-\tau_1) + \sum_{\tau_1=0}^{M-1}\sum_{\tau_2=0}^{M-1} h_2(\tau_1, \tau_2)x(n-\tau_1)x(n-\tau_2) + \]
\[ \sum_{\tau_1=0}^{M-1} \sum_{\tau_2=0}^{M-1} \sum_{\tau_3=0}^{M-1} h_3(\tau_1, \tau_2, \tau_3) x(n-\tau_1)x(n-\tau_2) x(n-\tau_3) + ...\]
Several optimization can be made with this form.
\subsection{Theory of the inverse model}
\begin{figure}[h!]\centering
\includegraphics[scale=.4]{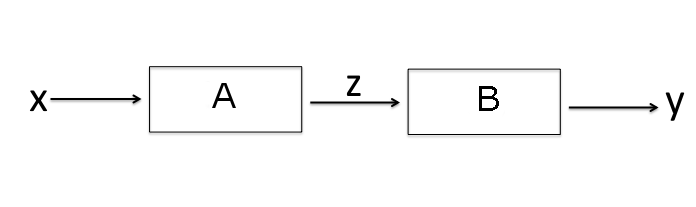} 
\caption{Series interconnection} \label{fig:series}
\end{figure}

\begin{figure}[h]\centering
\includegraphics[scale=.4]{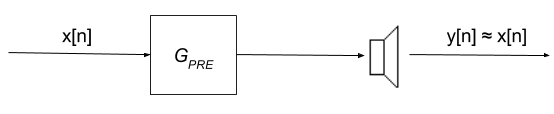} 
\caption{Method to correct the signal with a pre-inverse $G_{PRE}$} \label{fig:preinverse}
\end{figure}

In order to correct the distortion, it has been fundamental create a model able to simulate the loudspeaker behavior; specially to understand the problems related to the memory requirements.\\
After this step an inverted filter must be implemented. A nonlinear inversion introduces several problems.\\
In this section, it will be presented the notations and the mathematical justifications used to invert the model.\\\\
In this chapter, we will use these common notations:
\begin{itemize}
\item $H$ represents the system under investigation.
\item $H^p$ will be the $p$-th order term of the system $H$, similarly for the inverse system $G^p$
\item $H_M$ is the Volterra model of the system considered, it contains the first three orders of the Volterra expansion.
\item $G$ will be used to identify the inverse of the loudspeaker system.
\item $G_M$: We will use this notation to identify the Volterra model of $G$. In the same way of $H_M$, it will be considered the first three orders of the truncated series.
\end{itemize}
In order to define the concept of inversion, we will use the operation called series interconnection, it can be given between two systems:
\paragraph{series interconnection} Let $A$, $B$ two systems, a series interconnection is the result of an input $x$ into $A$ which results in an output $z$ that is in turn the input for the system $B$ which results in an output $y$. As shown in fig. \ref{fig:series}.
Formally:
\[ z = A(x)\]
\[ y= B(z) \]
\[ y = B(A(x))\]
A particular attention should be put in case of nonlinear systems. In this case in general, $B(A(x))$ is different from $A(B(x))$.\\\\
Now we can define the inversion of a nonlinear system with memory. It will be used for the expansion of a Volterra system, as in \cite{designvolterra}:
\paragraph{inversion} A $p$-th order inverse $G$ of a given nonlinear system $H$ is a system that when connected in series with $H$ results in a system in which the first order Volterra kernel is the unity system and the second through $p$-th order Volterra kernel are zero.\\
%So, let $F$ a Volterra model composed by $H$ and $G$ in series, then:
Formally:
\begin{equation}
G(H(x(n))) = x(n) + \sum_{m=p+1}^{\infty} G^m(H^m(x(n)))
\end{equation}
Considering the inversion definition, we will call:
\begin{itemize}
\item \textit{Pre-inverse}: The inverse system $G_{PRE}$ is positioned before $H$ in the interconnection.  $H(G_{PRE}(x))=x$
\item \textit{Post-inverse}: $G_{POST}$ is put after $H$. $G_{POST}(H(x))=x$
\end{itemize}
The only way to correct the loudspeaker audio consist in a filter that adjust the signal before arriving to the system, as shown in \ref{fig:preinverse}. Then, a pre-inverse must be computed.
Now, we will report an important theorem related to $p-th$ order inverse of a Voltrra series \cite{preinverse}, it is crucial to justificate the method used to find the kernel of inverse model.\\\\
\textbf{Theorem}\\
A $p$-th order pre-inverse of a Volterra system is identical to its $p$-th order post-inverse.\\
\textbf{Proof}\\
Let $H$ a system characterized by the Volterra series and its post-inverse $G_{POST}^p$, with the notation $G_{POST_i}^p$ for the $i$-th order operator of $G_{POST}^p$.\\
\textbf{Idea}\\
Consider the first $p$ orders of $H$, denoted with $H^p$, we can write:
\[H^p G_{POST}^p = I \]
It can be rewritten as:
\[H^p I G_{POST}^p = I \]
It can also be expressed as:
\[ H^p G_{POST}^p H^p G_{POST}^p = I \]
Then, $G_{POST}^p H^p$ must be equal to $I \Rightarrow G_{POST}^p$ is equal to the pre-inverse.\\
In this idea it is not considered the terms above the order $p$. Below the complete proof is reported as in \cite{preinverse}:\\\\

\begin{figure}[h]\centering
\includegraphics[scale=.4]{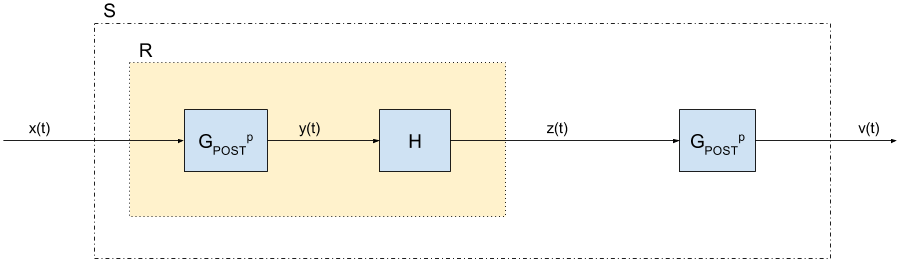} 
\caption{System $S$ composed by $R$ in series with $G_{POST}^p$} \label{fig:Dim_S_1}
\end{figure}

\begin{figure}[h]\centering
\includegraphics[scale=.4]{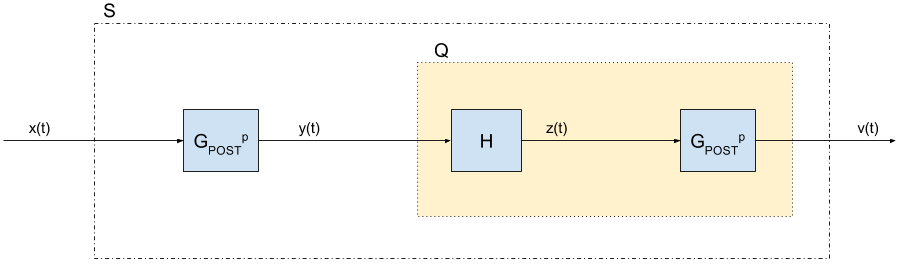} 
\caption{System $S$ composed by $G_{POST}^p$ in series with $Q$ } \label{fig:Dim_S_2}
\end{figure}

Consider a system $S$  defined as in fig \ref{fig:Dim_S_1}.
We call $R$ the sub-system composed in series by $G_{POST}^p$ and $H$. The same system $S$ can be alternatively seen as in fig. \ref{fig:Dim_S_2}.
Where $Q$ is the sub-system composed in series by $H$ and $G_{POST}^p$.\\
The Volterra series of system $R$ is given by:
\begin{equation}
z(t) = \sum_{n=1}^\infty R_n (x (t))
\end{equation}

If we consider the system $S$ as in fig. \ref{fig:Dim_S_2}, we can write
\begin{equation}
v(t) = Q(y(t)) = y(t) + \sum_{n=p+1}^\infty Q_n (y(t))
\end{equation}
expanding $y(t)$ as
\begin{equation}
y(t) = G_{POST}^p (x(t)) = \sum_{n=1}^p G_{POST_n}^p (x(t))
\end{equation}
we obtain
\begin{equation}
v(t) = \sum_{n=1}^p G_{POST_n}^p (x(t)) + \sum_{n=p+1}^\infty Q_n (y(t))
\label{eq:proof_5}
\end{equation}
From fig. \ref{fig:Dim_S_2} and the eq. \ref{eq:proof_5}
\begin{equation}
S_n(x(t)) = G_{POST_n}^p (x(t)) \text{ for } n \leq p
\label{eq:proof_6}
\end{equation}
Consider now the system $S$ defined as in fig. \ref{fig:Dim_S_1}, the following equation can be written:
\begin{equation}
z_n(t) = R_n(x(t))
\label{eq:proof_7}
\end{equation}
Besides, from definition of $S$ as in fig. \ref{fig:Dim_S_1}
\begin{equation}
S_1 (x(t)) = G_{POST_1}^p (z_1(t))
\label{eq:proof_8}
\end{equation}
using the eq. \ref{eq:proof_6}
\begin{equation}
G_{POST_1}^p(x(t)) = G_{POST_1}^p (z_1(t))
\end{equation}
$G_1$ is a linear operator, then it can be obtained
\begin{equation}
x(t) = z_1(t)
\label{eq:proof_9}
\end{equation}
Using equation \ref{eq:proof_7} and \ref{eq:proof_9}
\begin{equation}
R_1 (x(t)) = x(t)
\end{equation}
Now the induction will be used for the second part.\\
\textbf{n=2}\\
Considering $S_2$ from fig. \ref{fig:Dim_S_1}, we can write $S_2$ as
\begin{equation}
S_2(x(t)) = G_{POST_1}^p (z_2 (t)) + G_{POST_2}^p(z_1(t))
\label{eq:proof_10}
\end{equation}
replacing the equations \ref{eq:proof_6} and \ref{eq:proof_9} in \ref{eq:proof_10}
\[G_{POST_2}^p(z_1(t)) =G_{POST_1}^p (z_2(t)) + G_{POST_2}^p( z_1 (t))\]
Then
\begin{equation}
G_{POST_1}^p (z_2 (t)) = 0
\label{eq:proof_11}
\end{equation}
$G_{POST_1}^p$ is a linear operator, then
\begin{equation}
z_2(t) = 0
\label{eq:proof_12}
\end{equation}
Considering eq. \ref{eq:proof_7}:
\begin{equation}
R_2(x(t))=0
\end{equation}
Assuming $z_m (t) = R_n (x(t)) = 0,$ $n=2,....,m-1$ for $m\leq p$. We must prove $z_m(t) = 0$.\\
For the definition of $S$ we have the following equation
\begin{equation}
S_m(x(t)) = G_{POST_1}^p(z_m(t)) + G_{POST_2}^p (z_{m-1} (t)) + ... + G_{POST_m}^p (z_1 (t))
\label{eq:proof_13}
\end{equation} 
Using the induction hypothesis in order to write the eq. \ref{eq:proof_13}
\begin{equation}
S_m (x(t)) = G_{POST_1}^p (z_m (t)) + G_{POST_m}^p (z_1(t))
\end{equation}
Using eq. \ref{eq:proof_6} and \ref{eq:proof_9} in the previous equation, it can be obtained
\[ G_{POST_m}^p (z_1(t)) = G_{POST_1}^p (z_m (t)) + G_{POST_m}^p (z_1 (t))  \]
Then
\begin{equation}
G_{POST_1}^p ( z_m (t)) = 0
\end{equation} 
From the previous example and the linearity of $G_{POST_1}^p$
\begin{equation}
z_m(t) = 0
\end{equation}
Finally replacing with eq. \ref{eq:proof_7}:
\begin{equation}
R_m(x(t))=0
\end{equation}
This means that using $G_{POST}^p$ before the system $H$ we obtain
\begin{equation}
R_n (x(t)) = \begin{cases}
               x(t) \text{     if } n=1\\
               0 \text {         if } n=2,...,p
            \end{cases}
\label{eq: proof_end}
\end{equation}
This is the consequence of a pre-inverse application, then $G_{POST}^p$ is equal to $G_{PRE}^p$.
\begin{flushright}
$\blacksquare$
\end{flushright}

In short using a Volterra series, it can be found both pre-inverse and post-inverse, without any difference.\\\\
There are three main approaches to estimate an inverse of a system $H$ \cite{inverse}:
\begin{itemize}
\item In a first step, the forward model $H_M$ is estimated with an optimization algorithm, using input data $x$ and output data $y$. In step two, $H_M$ is inverted analytically.
\item The forward model $H_M$ is estimated as the previous method, this model is used in series with an inverse model $G_M$ and the inverse model parameters are estimated in this setting, by minimizing the difference between the input $x$ and the simulated output $y_{H_M}$.
\item The identification is done in one step by identifying the inverse $G_M$ directly, using input $y$ and output data $x$.\\ This approach assumes that the pre-inverse and the post-inverse are interchangeable.
\end{itemize}
In this project the third method is used because the result is generally better than for the first method \cite{inverse_estimating}. Besides, In pre-distortion applications, the third approach is more commonly used than the second, because it seems to perform slightly better \cite{comparisonlearning}.
\paragraph{computational cost} The first order Volterra filter requires $M$ steps to be computed, it is a discrete convolution between two vectors of this dimension.\\ Without any optimization, the method need a matrix with dimension $M^2$ for the second order. For the third order a tridimensional matrix must be used, this imply that for the third order Volterra filter $O(M^3)$ steps are required.\\ 
In general, let $f$ the Volterra filter algorithm of order $i$, then $f \in O(M^i)$. Due to the complexity of the model, a particular attention should be put on the choice of $M$.\\\\
Considering the performance required in audio applications, we focused the study on the second and third order Volterra filters. So, every consideration about filters and implementation will be assumed related to the truncated series:
\[
y(n) = h_0 + \sum_{\tau_1=0}^{M-1} h_1(\tau_1)x(n-\tau_1) + \sum_{\tau_1=0}^{M-1}\sum_{\tau_2=0}^{M-1} h_2(\tau_1, \tau_2)x(n-\tau_1)x(n-\tau_2) + \]
\begin{equation}
\sum_{\tau_1=0}^{M-1} \sum_{\tau_2=0}^{M-1} \sum_{\tau_3=0}^{M-1} h_3(\tau_1, \tau_2, \tau_3) x(n-\tau_1)x(n-\tau_2) x(n-\tau_3)
\label{eq:volterra_3_order}
\end{equation}

\section{Kernel Estimation}
\begin{figure}[h]\centering
\includegraphics[scale=.4]{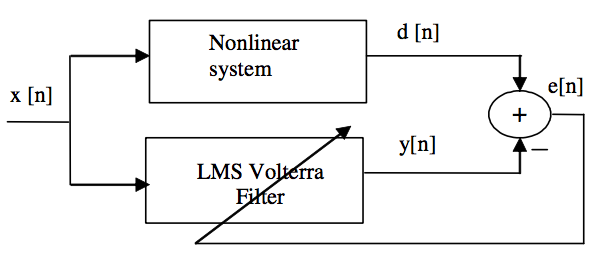} 
\caption{Method used to estimate the kernel} \label{fig:lmsadaptive}
\end{figure}

Before the filter creation, we need that eq. (\ref{eq:volterra_3_order}) replicates the loudspeaker behavior. The kernel must be estimated on the input data and its related output data.\\\\ 
The algorithm is a variation of the well known Least Mean Square (LMS), called Normalized Least Mean Square (NLMS), it will be discussed in the following. It is becoming a standard method for nonlinear system characterization based on Volterra series.\\\\
A set of typical input signals is chosen, each input signal $x(n)$ must to be sent at loudspeaker and its output is collected with a measurement system. The loudspeaker output represents the desired signal of the model, so it will be called $d(n)$.\\
The set of input and desired signals represent the knowledge base of the loudspeaker, it will be used to extract the information about the system and build the model. If this set is not representative, the model will not be reliable. To avoid this situation different signals have been used in our implementation, the details will be presented in the next chapter.\\
The previous input $x(n)$ is processed with the filter in order to obtain the output $y(n)$ related to current kernel. Now the correction step: the difference between $d(n)$ and $y(n)$ is computed and it is used to correct the filter, it represents the error and it is used as weight: if the error is large, the correction must be large.
This process is described in fig. \ref{fig:lmsadaptive}. \\
The estimation of the inverse model exploits the same strategy, inverting the order of input and desired signal. The desired signal $d(n)$ is used as input for the model,  $x(n)$ as the desired output signal.\\\\
Recalling the definition of the various types of inverse, the previous method allows to construct a post-inverse; because the model maps an output of a system to its initial input.\\
The employment of the same method during the pre-inverse research is justifiable thanks to the properties of the Volterra kernel, as seen before the pre-inverse and the post-inverse kernel are identical.
\subsection{Estimation Algorithm}
Vector operations can be used to evaluate the equation (\ref{eq:volterra_3_order}) in a convenient way. Fixed the memory $M$, the values related to $\prod_{j=1}^n x(t-\tau_j)$ can be pre-computed, for all orders.\\
For this operation three structures must be used, respectively with dimension $M$, $M^2$ and $M^3$, one for each order.\\\\
$x_1(k) = [ x_k; x_{k-1}; $ ... $; x_{k-M+1} ] $\\
$x_2(k) = x_1(k)^{T} * x_1(k)$\\
$x_3(k) = x_1(k)^{T} * x_2(k)$ \\\\
We implemented an optimization presented in \cite{lmsestimation}, this version exploits the symmetric kernel to avoid the computation of duplicate elements.
Besides, according to the same paper, the vectorial form is used to memorize the data related to the three orders.\\ 
In the second order, the matrix is put in a vector, placing side by side the rows. The third order is obtained collecting the $M$ matrices in the same vector.\\
This method is useful because the application of $i$-th order of the filter consists in a scalar product between $h_i^T$ and $x_i$.\\\\
The computation of the first order remains unchanged, the past $M$ values of the input signal is taken:\\
$x_1(k) = [ x_k, x_{k-1}, $ ... $ x_{k-M+1} ] $\\\\
The second order is computed considering only the lower triangular part of the matrix obtained from $x_1(k)^{T} * x_1(k)$, the result is a vector formed as:\\
$x_2(k) = [x^2(k), x(k)x(k-1),..., x(k)x(k-M+1),x^2(k-1),$
\begin{flushright}
$ x(k-1)x(k-2),..., x^2(k-M+1)]$\\
\end{flushright}
The third order follows the logic of the second order computation, for all $M$ matrices. The code used for this step is:\\
\begin{center}
\begin{minipage}{\linewidth}
\begin{lstlisting}[language=C++]
 int x2Start = 0;
 int x3Index = 0;
 for (i = 0; i<M; i++) {
	 // vectorSize contains the dimension of the input vectors
	 for (j = x2Start; j<vectorSize[1]; j++) {
		 x3[x3Index] = x1[i] * x2[j];
		 x3Index++;
	 }
	 x2Start = x2Start + M - i;
 }
\end{lstlisting}
\end{minipage}
\end{center}
the result is a vector with the following form:\\
$x_3(k) = [ x^3(k), x^2(k)x(k-1),..., x^2(k)x(k-M+1), x^3(k-1),$
\begin{flushright}
$ x^2(k-1)x(k-2),..., x^3(k-M+1)]$
\end{flushright}
With this method can be achieved a reduction of computational steps. The first order vector has dimension $M$, the second order vector $M*(M+1)/2$, the third order has dimension $M*(M+1)*(M+2)/6$.\\ 
This method will be useful also in the parallelization phase, because minimizes the memory transfer. However, the complexity of the algorithm doesn't change, the filter behaves as $O(M^3)$.\\\\
Now, NLMS algorithm can be applied following these steps:\\
\begin{algorithm}[H]
\SetAlgoLined
d = desiredSignal\;
h = initKernel()\;
\While{iteration < maxIterations}{
	\While{k < length(desiredSignal)}{
		$y_i(k) = h_i * x_i(k)^T, i=1,2,3$\;
		$y(k) = \sum_i y_i(k), i=1,2,3$\;
		$e(k) = d(k) - y(k) $\;
		$h_i = h_i + \mu_i e(k) x_i(k),  i=1,2,3$\;
	}
	iteration = iteration + 1\;
}
\caption{Normalized Least Mean Square algorithm}\label{alg:nlms}
\end{algorithm}
The Normalized Least Mean Square algorithm uses a normalized step-size $\mu_i$:
\[\mu_i = \frac{\alpha_i}{x_i(k)^T x_i(k) + \phi}\]
The constants $\alpha_i$ and $\phi$ are positive, and  $0<\alpha_i < 2$ and  $0<\phi < 1$ according to \cite{lmsestimation}.\\
The step-size parameter $\alpha_i$ controls the rate of convergence and stability.
The best approximation of the unknown system output is the identity, the initialization of the kernel vectors is done as follow:
\begin{itemize}
\item $h_1$: the first component equal to $1$, the others $0$
\item $h_2$, $h_3$: all components are fixed to $0$
\end{itemize}
\section{Implementation notes}
For data analysis and the creation of tests, the framework \textit{iPython} ( ipython.org ) has been used. It provides a powerful python shell with a support for interactive data visualization.\\\\
Regarding the estimation algorithm of Volterra kernel, currently it is not available an open-source library that provides an implementation. During this project, a library has been developed for this purpose.\\
The core of the library has been developed in \textit{C++}, considering the performance required and mathematical libraries, that will be useful for the future developments.\\\\
In order to simplify the exchange of data between the python interface and the C++ core, a common format has been used, called JSON ( JavaScript Object Notation ).
This method is widely used in practice.\\
JSON format is human readable and, at the same time, it easy to parse by a software. For these properties it became a standard method, currently all modern languages provide a library to parse and create objects with JSON format.\\
A typical estimation object, exchanged between interface and core, can be seen below:\\
\begin{center}
\begin{minipage}{\linewidth}
\begin{lstlisting}[language=json]
    { "alpha1": 1.0, 
    "alpha2": 0.4,
    "alpha3": 0.3, 
    "iterations": 3000, 
    "memory": 60, 
    "errorMax": 5.5e-05
    "input": [
        1.3197036988503929e-07, 
        0.00054189307967447112, 
        ... ],  
    "desired": [
        -5.828741642956409e-09, 
        -3.6815580229155927e-05, 
        ... ]   
	}
\end{lstlisting}
\end{minipage}
\end{center}
Two versions of the estimation method have been provided, the first implementation realizes exactly the algorithm logic of page 92; the second version uses more memory in order to reduce the computation time.\\
In the NLMS algorithm, the vector $x_i$ must be recomputed for every $k$, in all iterations; this operations always requires $O(M^3)$. It's possible pre-compute the vectors $x_i$ for the entire signal dimension.\\
This method is hardly applicable in context with big training signal, but it has been successful used in our tests, where signals have dimension of several thousands of points. The reduction of computation time compared with the standard implementation is on average of $13\%$.\\ 
The initial training step is an algorithm that falls back into offline processes, where the entire signal is available. This will not be possible with real-time filter, where only the latest part of the signal will be known.
\paragraph{Overfitting} The overfitting status occurs when a model achieves an excessive specialization on training data. It implies the inability to generalize the system behavior, and forecasting correctly the unknown input.\\\\
A significant improvement has been obtained with a threshold for the mean error, in terms of performance and result quality. The threshold is compared with the mean error, in every iteration. If this threshold becomes greater than the error, the estimation is interrupted.\\
This threshold allows to stop the process when the kernel have a sufficiently good approximation, in general less than 100 iterations are needed.
\subsection{Prototype for GPU}
In order to improve the performance in future, we also implemented a different core prototype; it exploits the GPU ( Graphics Processing Unit ). This hardware is specialized to high-performance computing with matrix operations. So, applications based on this logic can benefit of this implementation choice.\\
The prototype has been implemented with CUDA (Compute Unified Device Architecture), developed by NVIDIA. It's based on a massive parallelism architecture, integrated with a general purpose language. A powerful extension of the framework is available in C/C++.\\\\
Currently, CUDA has been used to parallelize the operation $h_i^T x_i$; with the mathematical library cuBLAS ( CUDA Basic Linear Algebra Subroutines ).\\
Unfortunately, a simple multiplication of the vectors made by GPU is not useful due to high memory latency. In our test system, the GPU is a dedicated hardware with its own memory. It is necessary to transfer the working variables to its memory space.\\
The transfer time represents the bottleneck of the computation due the dimension of the working vectors. This method will be changed exploiting a pipeline to hide the memory latency.

\part{Results and Conclusion}
\chapter{Measurements and Test}
\begin{figure}[h]\centering
\includegraphics[scale=.2]{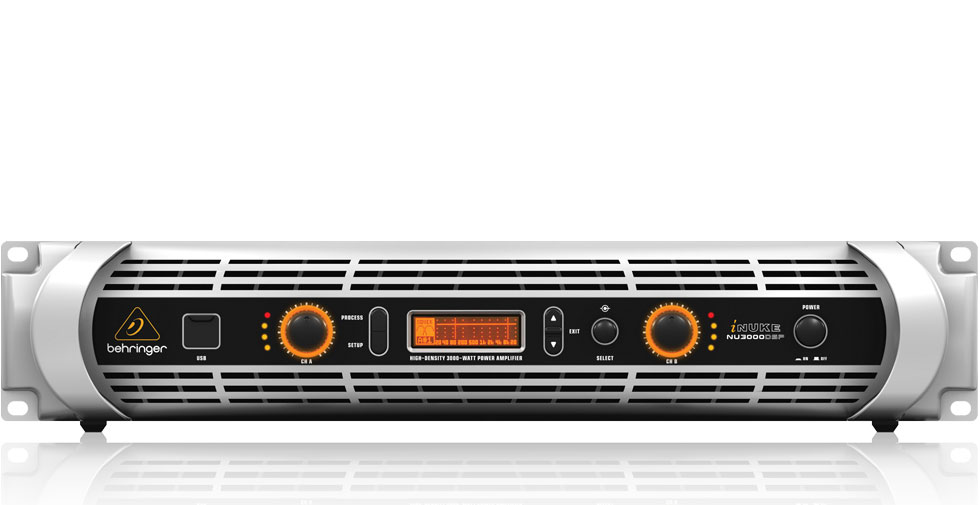}
\caption{Amplifier used for measurements.} \label{fig:loud-amplifier}
\end{figure}

The loudspeaker measurements have been made in collaboration with the Institute of Applied Mathematics and Physics, at the Zurich University of Applied Sciences.\\\\
All filters have been tested considering a sub-woofer, which as previously presented, produces the most significant distortions.\\
The selected input signals are distributed in a low frequency region, in particular:
\begin{itemize}
\item Monochromatic signals from 20 Hz to 150 Hz
\item Signals obtained as a superposition of sine wave, with the distance in frequencies of 2, 3, 5, 6 Hz. In this section they are referred to as \textit{multisine2}, \textit{multisine3}, \textit{multisine5}, \textit{multisine6}.
\item White noise
\item Chirp, a signal in which the frequency increases (or decreases) with the time
\end{itemize}
The sub-woofer used in the investigation is a SONY Box Subwoofer XS- NW1202E, with a power amplifier Nuke NU3000 High density 3000W. The equipment is displayed in fig \ref{fig:loud-amplifier}.\\\\
\begin{figure}[h]\centering
\includegraphics[scale=.2]{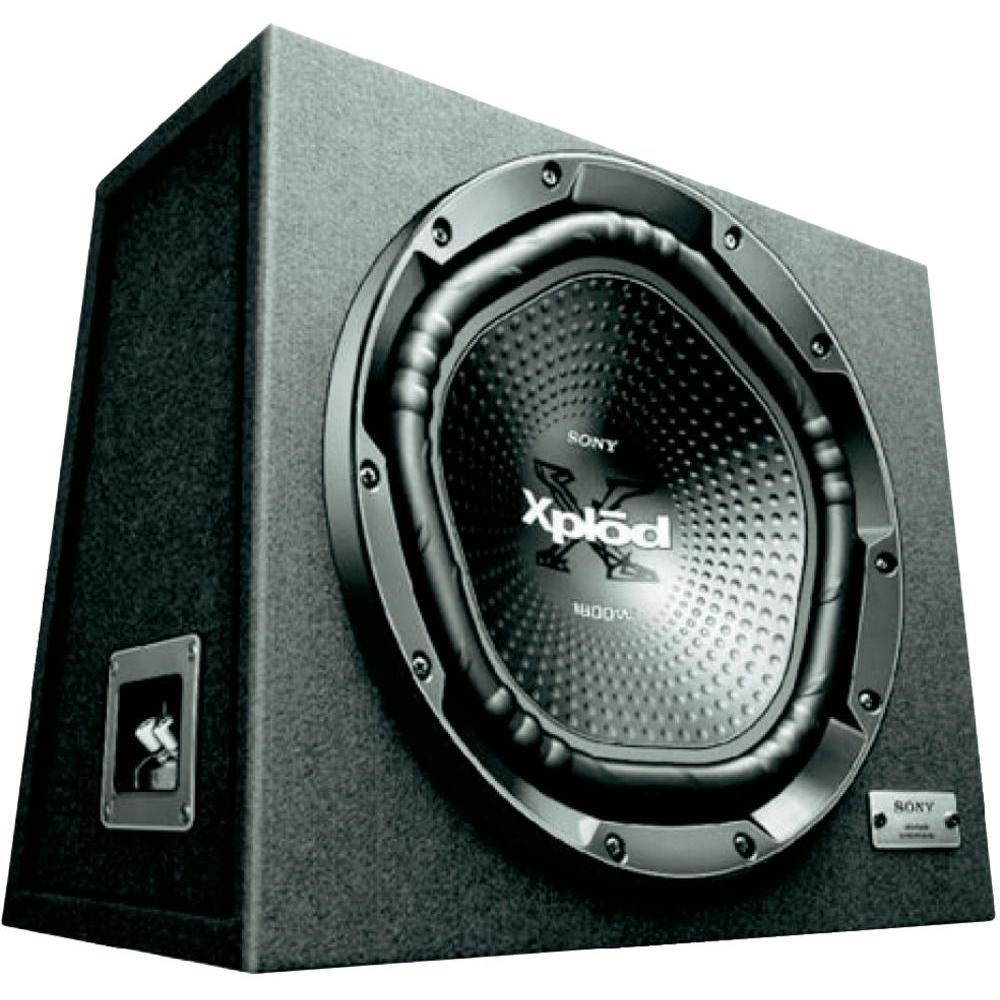}
\caption{Loudspeaker used for measurements.}
\end{figure}
All measurements have been performed with a standard technique called \textit{interferometry}, which is explained in the following section.\\
A complete description of the measurement phase can be found in \cite{measurementsreport}.

\subsection{Interferometry}
\begin{figure}[h]
\begin{minipage}{.5\textwidth}
\caption{Michelson configuration}\label{fig:standardinf}
\includegraphics[scale=.5]{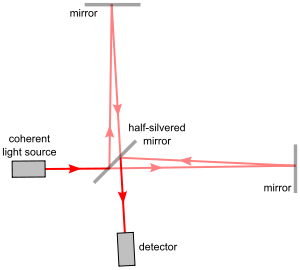} 
\end{minipage}
\begin{minipage}{.5\textwidth}
\caption{Our configuration}\label{fig:ourinf}
\includegraphics[scale=.5]{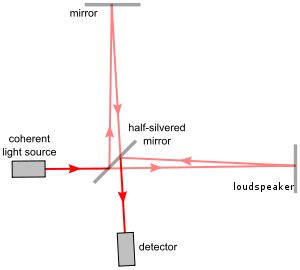} 
\end{minipage}
\end{figure}

In this technique a laser is used for a monochromatic light beam, which is places at one of the four position described in the fig \ref{fig:standardinf}. A single incoming beam is divided into two identical beams by a partially reflecting mirror, which is positioned in the middle of the experiment space. Each beam follows a different path; at the end of the path there is another mirror that reflects the beam in the inverse path. At this point they are recombined before they arrive at the detector.\\\\
The path difference creates a phase difference between the signals, this difference can be analyzed to obtain the characterization of the path.
In this case, a reflecting coating is deposited in the loudspeaker membrane, it has been positioned at the end of a path. The interference pattern about the membrane movements can be collected.\\
This configuration is described by the figure \ref{fig:ourinf}.\\\\
This method has been very useful because the movement of the membrane represents the sound that is produced by the loudspeaker.\\
Thanks to this methodology, the model built is based on high reliable data.

\section{Test}
\begin{figure}[H]
\caption{Comparison between the original signal and decimated signal.}\label{fig:comparison}
\begin{minipage}{.5\textwidth}
\centering
\includegraphics[scale=.32]{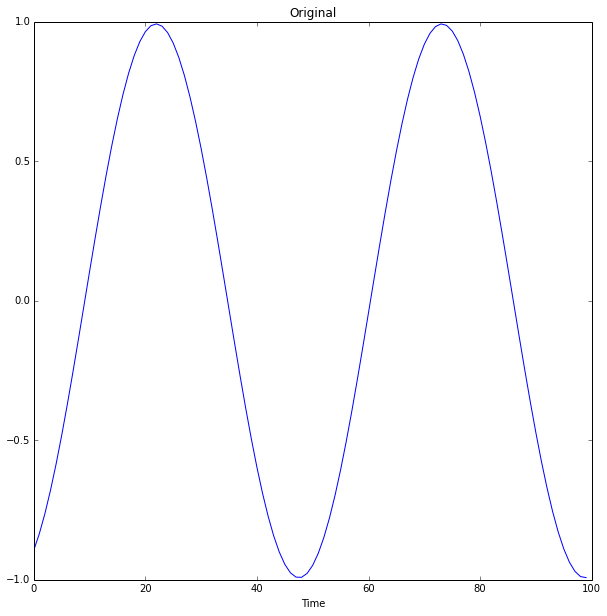} 
\end{minipage}
\begin{minipage}{.5\textwidth}
\centering
\includegraphics[scale=.32]{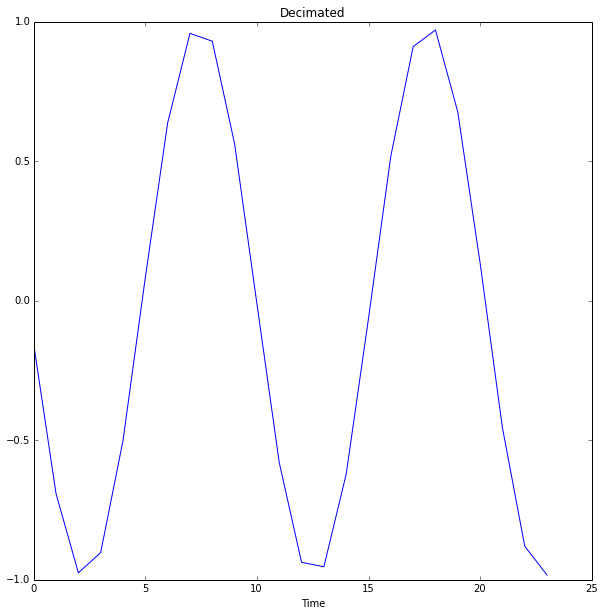} 
\end{minipage}
\end{figure}

In this section will be presented a set of test to summarize the results achieved during this work. To evaluate correctly the generalization capability of the filters, 
they were tested using signals different from that used in estimation step. Each signal has been divided into two sub-signals: the \textit{training signal} composed by the $70\%$ of the original signal; the other part is used for the \textit{test signal}.\\\\
In order to compare the filters with a common metric, for each one is reported the \textit{Mean Squared Error} (MSE) related to a set of tests:
\begin{itemize}
\item Chirp
\item $20Hz$
\item $50Hz$
\item $70Hz$
\item $multisine6$
\item $multisine3$
\end{itemize}
The chirp is useful because the frequencies are distributed for the entire interested spectrum. The monochromatic signals were used because their easiness to show the harmonics introduced by the system. Superimposed signals have been used because they imply a more complex behavior of the loudspeaker.\\\\
All signals obtained from the measurements are sampled at the frequency of $2560Hz$. To reduce the complexity of the problem, a downsampling has been applied with a decimation factor of $5$, so the resulting signals have a sampling frequency of $512Hz$. The difference between the original signal obtained from the measurement and the downsampled signal can be seen in \ref{fig:comparison}.
The decimation factor has been chosen in according to the constraint of the Shannon sampling theorem, this value does not introduce aliasing because the highest frequency is $150Hz$, then the new sampling frequency continue to be greater than Nyquist frequency.\\\\
At first, the results of the loudspeaker model will be presented. This step has been fundamental to understand the model features and the estimation parameters; in particular the memory requirements and the learning coefficients $\alpha_i$, $\phi$ (algorithm \ref{alg:nlms}).\\
In the second part the inverse filter results will be presented with a special attention to the error correction difference between a linear and nonlinear filters. % use \myChapter command instead of \chapter
	\include{test}
	\chapter{Conclusion and future development}
\begin{figure}[h]\centering
\includegraphics[scale=.2]{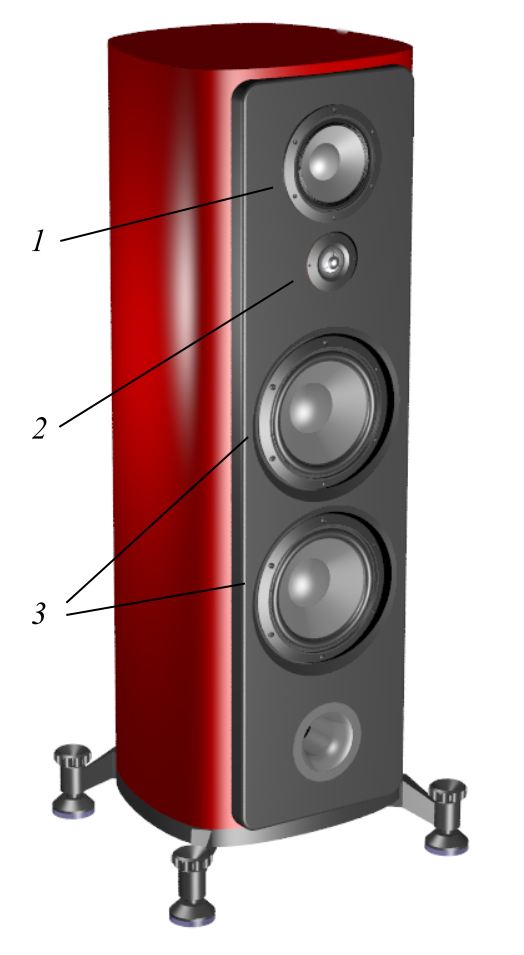} 
\caption{Loudspeaker system with different membranes, each one has a dedicated range of frequencies. 1 - Midrange (250Hz to 2 KHz), 2 - Tweeter (2KHz to 20KHz), 3 - Woofers (20Hz to 250Hz)} \label{fig:loudspekerscheme}
\end{figure}

\begin{figure}[h]\centering
\includegraphics[scale=.4]{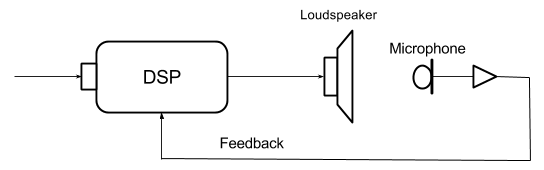} 
\caption{Feedback control}\label{fig:feedbackchannel}
\end{figure}

As seen in the results presented in the previous chapter, exploiting the Volterra series can correct a large amount of distortion produced by loudspeaker. In the last years there has been a renewed interest in the commercial and scientific context about the implementation of these filters; a lot of results presented in other papers consider only the first two orders of the series.\\
In this thesis a systematic set of tests has been made to find the best parameters and training methods, in order to estimate the Volterra kernel with high quality results on different types of signal. All tests have been performed considering the first three orders of the series, therefore we provided a practical contribution to the adaptive nonlinear filters based on this series.\\\\
All scientific papers found about Volterra filters for loudspeaker signal correction are based on signals recorded with a microphone, often subject to non negligible noise. In this project, all filters exploited signals collected with an interferometer from the Zurich University of Applied Sciences. This method is very accurate accurate and the results can be considered more reliable than those collected with a microphone.\\\\
We have written a paper summarizing the results of the loudspeaker model, it must be submitted to the journal (\textit{Nonlinear Volterra model of a loudspeaker behavior based on interferometry measurements. Alessandro Loriga, Elizabeth Dumont}).\\\\
A set of systematic tests has been performed in order to ensure a complete vision of the model requirements and parameters. The evaluation has been made for different types of signals (monochromatic, superimposed ...) and the behavior of the filters are clear in several situations.\\\\
All algorithms have been implemented in a C++ library, composed of two main parts: \textit{Handler} and \textit{Estimator}. \\
The estimation algorithm can be called from other methods allocating a Handler instance. Different implementations of the estimation process can be written with a derived class that inherits from the Estimator class. This structure allows the library to be easily expanded with new features.\\\\
Future developments of this project will be made by \textit{Florence Technologies s.r.l} and \textit{Intranet Standard GmbH}, the aim of which consists of a filter implementation for dedicated loudspeakers; Zurich University of Applied Sciences will carry out further measurement phases. In the following paragraphs we will introduce the main steps of the project.\\\\
A good loudspeaker system is composed of a set of different membranes, as shown in fig. \ref{fig:loudspekerscheme}. In this project we focused on a sub-woofer, a similar study can be made for woofers of mid-ranges.\\\\
The changing of membrane characteristics during the loudspeaker life-cycle can not be neglected: elasticity, movement capabilities and system resonance do not remain the same due to loudspeaker usage, therefore the filters can not be considered reliable after a certain period.\\
Our idea is to provide a feedback channel obtained with a good quality microphone. This channel is a standard method in control systems and it is described in the fig. \ref{fig:feedbackchannel}; this is the same method presented in the Active Nose Control. \\
We think that this periodical calibration will not require a large number of iterations because the initial kernel will be a good estimation of the searched kernel. This implementation step has to pass for the optimization of kernel estimation algorithm.\\\\ 
A code optimization must be performed for the filter application, as each filter has a natural delay due to the computational time required. In an audio context, it should be very small, but in live music context this is a crucial point. It is impossible play an instrument with a system that introduces a delay greater than a certain threshold.\\
In all likelihood the entire computation system will be implemented in a GPU-accelerated embedded system. In these systems the GPU is integrated in the main board, reducing the transfer latency between the CPU memory and the device memory.\\\\
The last step will consider a method to extend the filters from a single loudspeaker to multiple devices. This step creates several communication problems between the various parts that are essential for the synchronization.\\
The communication between the components will probably be implemented with Ethernet or wi-fi, because several low latency synchronized protocols exist for these connections.
		
%********************************************************************
% Back matter
%********************************************************************
	\bibliographystyle{alpha}
	\bibliography{tesi}

\end{document}